\documentclass[%
 aip,
 pop,
 amsmath,amssymb,
 reprint,%
]{revtex4-1}

\usepackage{graphicx}
\usepackage{dcolumn}
\usepackage{bm}

\usepackage[utf8]{inputenc}
\usepackage[T1]{fontenc}
\usepackage{mathptmx}

\begin{document}

\preprint{AIP/123-QED}

\title[Recoil Effects on Reflection from Relativistic Mirrors in Laser Plasmas]{Recoil Effects on Reflection from Relativistic Mirrors in Laser Plasmas}

\author{P.~Valenta}
\email{petr.valenta@eli-beams.eu}
\affiliation{ELI Beamlines, Institute of Physics, Czech Academy of Sciences, Na Slovance 2, 18221 Prague, Czech Republic} 
\affiliation{Faculty of Nuclear Sciences and Physical Engineering, Czech Technical University in Prague, Brehova 7, 11519 Prague, Czech Republic}%
	
\author{T.~Zh.~Esirkepov}%
\affiliation{Kansai Photon Science Institute, National Institutes for Quantum and Radiological Science and Technology, Umemidai 8-1-7, Kizugawa, 619-0215 Kyoto, Japan}%

\author{J.~K.~Koga}%
\affiliation{Kansai Photon Science Institute, National Institutes for Quantum and Radiological Science and Technology, Umemidai 8-1-7, Kizugawa, 619-0215 Kyoto, Japan}%

\author{A.~S.~Pirozhkov}%
\affiliation{Kansai Photon Science Institute, National Institutes for Quantum and Radiological Science and Technology, Umemidai 8-1-7, Kizugawa, 619-0215 Kyoto, Japan}%

\author{M.~Kando}%
\affiliation{Kansai Photon Science Institute, National Institutes for Quantum and Radiological Science and Technology, Umemidai 8-1-7, Kizugawa, 619-0215 Kyoto, Japan}%

\author{T.~Kawachi}%
\affiliation{Kansai Photon Science Institute, National Institutes for Quantum and Radiological Science and Technology, Umemidai 8-1-7, Kizugawa, 619-0215 Kyoto, Japan}%

\author{Y.-K.~Liu}%
\affiliation{Leung Center for Cosmology and Particle Astrophysics, National Taiwan University, No. 1, Sec. 4, Roosevelt Rd., 10617 Taipei, Taiwan, R.O.C.}%

\author{P.~Fang}%
\affiliation{Leung Center for Cosmology and Particle Astrophysics, National Taiwan University, No. 1, Sec. 4, Roosevelt Rd., 10617 Taipei, Taiwan, R.O.C.}%

\author{P.~Chen}%
\affiliation{Leung Center for Cosmology and Particle Astrophysics, National Taiwan University, No. 1, Sec. 4, Roosevelt Rd., 10617 Taipei, Taiwan, R.O.C.}%

\author{J.~Mu}
\affiliation{ELI Beamlines, Institute of Physics, Czech Academy of Sciences, Na Slovance 2, 18221 Prague, Czech Republic}%

\author{G.~Korn}
\affiliation{ELI Beamlines, Institute of Physics, Czech Academy of Sciences, Na Slovance 2, 18221 Prague, Czech Republic}%

\author{O.~Klimo}
\affiliation{ELI Beamlines, Institute of Physics, Czech Academy of Sciences, Na Slovance 2, 18221 Prague, Czech Republic}%
\affiliation{Faculty of Nuclear Sciences and Physical Engineering, Czech Technical University in Prague, Brehova 7, 11519 Prague, Czech Republic}%

\author{S.~V.~Bulanov}
\affiliation{ELI Beamlines, Institute of Physics, Czech Academy of Sciences, Na Slovance 2, 18221 Prague, Czech Republic}%
\affiliation{Kansai Photon Science Institute, National Institutes for Quantum and Radiological Science and Technology, Umemidai 8-1-7, Kizugawa, 619-0215 Kyoto, Japan}%
\affiliation{Prokhorov Institute of General Physics, Russian Academy of Sciences, Vavilova 38, 119991 Moscow, Russia}%

\date{\today}

\begin{abstract}
Relativistic mirrors can be realized with strongly nonlinear Langmuir waves excited by intense laser pulses in underdense plasma. On reflection from the relativistic mirror the incident light affects the mirror motion. The corresponding recoil effects are investigated analytically and with particle-in-cell simulations. It is found that if the fluence of the incident electromagnetic wave exceeds a certain threshold, the relativistic mirror undergoes a significant back reaction and splits into multiple electron layers. The reflection coefficient of the relativistic mirror as well as the factors of electric field amplification and frequency upshift of the electromagnetic wave are obtained.
	
\end{abstract}

\maketitle



\section{\label{sec:introduction} Introduction}

A relativistic mirror may be defined as an object that reflects incoming radiation while moving at relativistic velocity. The theory of light reflection from such an object propagating in vacuum at arbitrary (subluminal) velocity was first formulated by Einstein in 1905 \cite{Einstein1905}. Since then, relativistic mirrors have been studied in many different contexts because of their great potential for both fundamental science and practical applications.

An electromagnetic wave incident on a relativistic mirror undergoes energy and frequency change due to the double Doppler effect. In a co-propagating configuration in the laboratory frame of reference, the reflected wave is stretched, its amplitude is lowered and its frequency is downshifted. On the contrary, in a counter-propagating configuration, the reflected wave is compressed, amplified and its frequency is upshifted.  

Relativistic mirrors can be realized by irradiating plasma targets with intense laser pulses (see Ref.~\onlinecite{Bulanov2013} for a review and the literature cited therein). They appear in laser plasma as thin dense electron (or electron-ion) shells accelerated to relativistic velocities. Various schemes that lead to the generation of relativistic mirrors have been described in theoretical as well as experimental studies (e.g. double-sided mirror \cite{Kulagin2007, Kulagin2007a, Esirkepov2009, Meyer-ter-Vehn2009, Bulanov2010, Wu2010, Wu2011, Wu2012, Andreev2013, Kiefer2013, Ma2014}, oscillating mirror \cite{Bulanov1994, Lichters1996, Naumova2004, Baeva2006, Wheeler2012, Vincenti2019}, sliding mirror \cite{Pirozhkov2006, Pirozhkov2007a}, flying mirror realized with strongly nonlinear Langmuir waves \cite{Bulanov2003, Kando2007, Pirozhkov2007, Kando2009, Lobet2013, Koga2018, Moghadasin2019} or electron density singularities \cite{Mu2019}) and, hence, have already proven the feasibility of this concept.

Nowadays, relativistic mirrors in plasmas are actively studied as a unique tool for fundamental research (e.g. light intensification towards the Schwinger limit \cite{Bulanov2003}, investigation of photon-photon and Delbr{\"u}ck scattering \cite{Koga2012, Koga2018}, analog black hole to investigate Hawking radiation and the information loss paradox \cite{Chen2017}) and for many practical applications in diverse fields; depending on whether the configuration is co-propagating or counter-propagating in the laboratory frame of reference, relativistic mirrors might be used either for acceleration of ions (e.g. for hadron therapy \cite{Bulanov2002}) or for producing coherent high-brightness radiation with wavelengths ranging from x-ray to gamma-ray (e.g. for molecular imaging \cite{Neutze2000}, attosecond spectroscopy \cite{Krausz2009}).

Maximization of the reflected radiation energy requires a more intense incident electromagnetic wave. However, sufficiently strong incident light can significantly affect the motion of the relativistic mirror (i.e. its radiation pressure can stop or destroy the mirror). In the present paper, we aim at a closer description of the recoil effects on a reflection from the relativistic mirror. We study the interaction of strongly nonlinear Langmuir waves with an incident counter-propagating electromagnetic wave as well as the properties of the reflected radiation. We discuss the regimes when the relativistic mirror undergoes a significant back reaction. We find the threshold of the onset of the recoil effects.

The paper is organized as follows: in section~\ref{sec:recoil} we derive the threshold for the energy of the incident electromagnetic wave, in section~\ref{sec:langmuir_wave} we discuss the physical realization of relativistic mirrors in laser plasma and in section~\ref{sec:pic_simulations} we demonstrate the results of one-dimensional (1D) particle-in-cell (PIC) simulations and compare them with the analytical calculations.

\section{\label{sec:recoil} Recoil effects on reflection from relativistic mirrors}

For a relativistic mirror propagating at constant velocity $ v_M $ in vacuum, the frequency upshift of a normally incident counter-propagating electromagnetic wave is given by \cite{Einstein1905}
\begin{equation}\label{eq:upshift_factor}
\frac{\omega}{\omega_0} = \frac{1 + \beta_M}{1 - \beta_M} = \frac{\gamma_{M} + \sqrt{\gamma_{M}^2 - 1}}{\gamma_{M} - \sqrt{\gamma_{M}^2 - 1}} \approx 4 \gamma_{M}^2,
\end{equation}
where $ \omega $ and $ \omega_0 $ are the frequency of the reflected and incident radiation, respectively, $ \beta_M = v_M / c $ is the velocity of the relativistic mirror normalized by the speed of light in vacuum $ c $ and $ \gamma_M = 1 / \sqrt{1 - \beta_M^2} $ is the corresponding relativistic Lorentz factor. The last term in Eq.~(\ref{eq:upshift_factor}) is obtained using the identity $ \gamma_{M} + \sqrt{\gamma_{M}^2 - 1} = (\gamma_{M} - \sqrt{\gamma_{M}^2 - 1})^{-1} $ and $ \gamma_{M} + \sqrt{\gamma_{M}^2 - 1} \approx 2 \gamma_{M} $ and is valid in the ultra-relativistic limit, i.e. when $ \gamma_{M} \gg 1 $. The factor of the electric field amplification of the reflected wave is given by \cite{Einstein1905}
\begin{equation}\label{eq:amplification_factor}
\frac{E}{E_0} = \frac{\omega}{\omega_0} \sqrt{R},
\end{equation}
where $ E $ and $ E_0 $ are the electric field of the reflected and incident radiation, respectively, and $ R $ stands for the reflection coefficient in terms of photon number.

The above Eqs.~(\ref{eq:upshift_factor}) and (\ref{eq:amplification_factor}) are derived in the approximation of a weak incident electromagnetic wave. Here, we analytically investigate the recoil effects of a counter-propagating electromagnetic wave normally incident on a relativistic mirror. This problem was briefly discussed in Ref.~\onlinecite{Pirozhkov2007a}. First, we consider the relativistic mirror in the form of an electron layer. We assume that all the electrons are characterized by the same momentum, the electromagnetic wave is monochromatic and the reflection coefficient in terms of photon number is equal to $ R $. The conservation of momentum and energy before and after the interaction can be then written as
\begin{eqnarray}
\label{eq:momentum_conservation}
\mathcal{N}_e p_{e0} - \mathcal{N}_\gamma p_{\gamma 0} = \mathcal{N}_e p_{e} + R \mathcal{N}_\gamma p_{\gamma} - \left( 1 - R \right) \mathcal{N}_\gamma p_{\gamma 0} ,\\
\label{eq:energy_conservation}
\mathcal{N}_e \mathcal{E}_{e0} + \mathcal{N}_\gamma \mathcal{E}_{\gamma 0} = \mathcal{N}_e \mathcal{E}_{e} + R \mathcal{N}_\gamma \mathcal{E}_{\gamma} + \left( 1 - R \right) \mathcal{N}_\gamma \mathcal{E}_{\gamma 0} .
\end{eqnarray}
Here, respectively, $ \mathcal{N}_e $ and $ \mathcal{N}_{\gamma} $ are the number of interacting electrons and photons. The subscript "$ 0 $" denotes the quantities before the interaction and the "$ - $" sign in Eq.~(\ref{eq:momentum_conservation}) denotes counter-propagating photons. The electron and photon momenta and energies can be expressed as
\begin{eqnarray}
\label{eq:momenta}
&p_e = m_e c \sqrt{\gamma_{e}^2 - 1}, \quad p_{\gamma} = \hbar \omega / c ,\\
\label{eq:energies}
&\mathcal{E}_e = m_e c^2 \gamma_e, \quad \mathcal{E}_{\gamma} = \hbar \omega ,
\end{eqnarray}
where the symbols $ \hbar $, $ \gamma_e $ and $ m_e $ denote the reduced Planck constant, the relativistic Lorentz factor of electrons and the electron rest mass, respectively. 

By combining Eqs.~(\ref{eq:momentum_conservation}) - (\ref{eq:energies}) we obtain the following formula:
\begin{eqnarray}\label{eq:omega}
&&\hbar \omega = \hbar \omega_0 \frac{\mathcal{N}_e \left( \mathcal{E}_{e0} + p_{e0} c \right)}{\mathcal{N}_e \left( \mathcal{E}_{e0} - p_{e0} c \right) + 2 R \mathcal{N}_\gamma \hbar \omega_0} \nonumber\\
&&= \hbar \omega_0 \frac{\mathcal{N}_e m_e c^2 \left( \gamma_{e0} + \sqrt{\gamma_{e0}^2 - 1} \right)}{\mathcal{N}_e m_e c^2 \left( \gamma_{e0} - \sqrt{\gamma_{e0}^2 - 1} \right) + 2 R \mathcal{N}_\gamma \hbar \omega_0}.
\end{eqnarray}
In the ultra-relativistic limit, i.e. when $ \gamma_{e0} \gg 1 $, Eq.~(\ref{eq:omega}) can be simplified as
\begin{equation}\label{eq:omega_approx}
\frac{\omega}{\omega_0} \approx 4 \gamma_{e0}^2 \frac{\frac{\mathcal{N}_e m_e c^2}{4 \gamma_{e0}}}{\frac{\mathcal{N}_e m_e c^2}{4 \gamma_{e0}} + R \mathcal{N}_\gamma \hbar \omega_0}.
\end{equation}
The two terms in the denominator of Eq.~(\ref{eq:omega_approx}) correspond to the energy of the electron layer and interacting photons, respectively. The resulting frequency upshift of the reflected radiation is determined by the relation between both terms:
\begin{subequations}
	\begin{eqnarray}
	&\omega/\omega_0& \ \approx \ 4 \gamma_{e0}^2 \quad \mathrm{for} \quad R \mathcal{N}_\gamma \hbar \omega_0 \ll \frac{\mathcal{N}_e m_e c^2}{4 \gamma_{e0}}, \label{eq:low_energy} \\
	&\omega/\omega_0& \ \approx \ \frac{\mathcal{N}_e m_e c^2 \gamma_{e0}}{R \mathcal{N}_\gamma \hbar \omega_0} \quad \mathrm{for} \quad R \mathcal{N}_\gamma \hbar \omega_0 \gg \frac{\mathcal{N}_e m_e c^2}{4 \gamma_{e0}}. \quad \label{eq:high_energy}	
	\end{eqnarray}
\end{subequations}
The limit~(\ref{eq:low_energy}) corresponds to the approximation of a weak incident electromagnetic wave, and produces the classical frequency upshift factor corresponding to the double Doppler effect (see Eq.~(\ref{eq:upshift_factor})). In the opposite limit~(\ref{eq:high_energy}), the incident radiation significantly affects the motion of relativistic mirror, so that the frequency of the reflected electromagnetic wave is in fact downshifted by the factor of $ \mathcal{N}_e m_e c^2 \gamma_{e0} / \left( R \mathcal{N}_\gamma \hbar \omega_0 \right)  \ll 1 $.

We define the threshold characterizing the recoil importance in this interaction as a midpoint between the limits given by Eqs.~(\ref{eq:low_energy}) and (\ref{eq:high_energy}), when the energy of the interacting photons is comparable to that of the electron layer,
\begin{equation}\label{eq:threshold}
R \mathcal{N}_\gamma \hbar \omega_0 = \varkappa \frac{\mathcal{N}_e m_e c^2}{4 \gamma_{e0}},
\end{equation}
where $ \varkappa < 1 $ is a small factor. Obviously, much less energy than the kinetic energy of the mirror can affect the reflection process.

\section{\label{sec:langmuir_wave} Relativistic mirror realized with a Langmuir wave}

A sufficiently short and intense laser pulse excites a strongly nonlinear Langmuir wave in underdense plasma \cite{Bulanov2016, Esarey2009}. The electron density modulations of the Langmuir wave in the wake of the laser pulse take the form of thin dense shells separated by cavities of length corresponding to the Langmuir wave wavelength $ \lambda_w $. A weak counter-propagating electromagnetic wave is partially reflected from these shells, undergoing energy and frequency change in accordance with the double Doppler effect. For this case, Eq.~(\ref{eq:upshift_factor}) becomes \cite{Bulanov2013}
\begin{eqnarray}\label{eq:upshift_factor_plasma}
&&\frac{\omega}{\omega_0} = \frac{1}{1 - \beta_w^2}\left(1 + \beta_w^2 + 2 \beta_w \sqrt{1 - \frac{\omega_{pe}^2}{\omega_0^2}} \right) \nonumber \\
&&= 2 \gamma_w^2 + 2 \gamma_w \sqrt{\gamma_w^2 - 1} \sqrt{1 - \frac{\omega_{pe}^2}{\omega_0^2}} - 1,
\end{eqnarray}
so that it includes the difference between the phase and group velocity in plasma. Here, $ \omega_{pe} = \sqrt{4 \pi n_e e^2 / m_e} $ is the Langmuir frequency corresponding to the background electron density $ n_e $, $ \beta_w $ is the phase velocity of the Langmuir wave normalized by $ c $ and $ \gamma_w = 1 / \sqrt{1 - \beta_w^2}$ is the corresponding relativistic Lorentz factor. The symbol $ e $ stands for the electron charge.

If the velocity of the electrons in the vicinity of the electron density spike exceeds the phase velocity of the Langmuir wave, i.e. $ \gamma_e > \gamma_w $, the Langmuir wave breaks. This corresponds to the Akhiezer-Polovin limit \cite{Akhiezer1956} for the longitudinal electric field, $ E_x $, of the Langmuir wave,
\begin{equation}\label{eq:akhiezer-polovin}
\frac{\max \left| E_x \right| e}{m_e \omega_{pe} c} > \sqrt{2 \left(\gamma_w - 1 \right)}.
\end{equation}
For the Langmuir wave at the threshold of wave-breaking, its reflection coefficient in terms of photon number, $ R $, is (see Ref.~\onlinecite{Bulanov2013})
\begin{equation}\label{eq:refl_coef}
R \approx \frac{\Gamma^2\left( 2 / 3 \right) }{2^2 \cdot 3^{4/3}}\left(\frac{\omega_{pe}}{\omega_0} \right)^{8/3} \frac{1}{\gamma_w^{4/3}},
\end{equation}
where $ \Gamma (x)$ is the Euler gamma function~\cite{Abramowitz1965}.

In order to estimate the threshold given by Eq.~(\ref{eq:threshold}) for the relativistic mirror realized with a breaking Langmuir wave, we represent, for simplicity, the incident laser pulse as an electromagnetic wavepacket with a rectangular profile and intensity $ I $, duration $ \tau $ and cross-sectional area $ S $. We assume normal incidence of this wavepacket on the relativistic mirror. The number of photons in the pulse, $ \mathcal{N}_\gamma $, is given by the following expression,
\begin{equation}\label{eq:n_photons}
\mathcal{N}_\gamma = \frac{I \tau S}{\hbar \omega_0}.
\end{equation}
For a nearly-breaking Langmuir wave, for which Eq.~(\ref{eq:refl_coef}) holds, approximately half of the plasma electrons are concentrated in the electron density spike in each wave period. Therefore, the number of interacting electrons, $ \mathcal{N}_e $, is
\begin{equation}\label{eq:n_electrons}
\mathcal{N}_e = \frac{n_e}{2} \lambda_w S.
\end{equation}

Using the reflection coefficient of the Langmuir wave of Eq.~(\ref{eq:refl_coef}), and, respectively, the number of interacting photons and electrons of Eqs.~(\ref{eq:n_photons}) and (\ref{eq:n_electrons}), we rewrite the threshold of Eq.~(\ref{eq:threshold}) in terms of the fluence (the product of intensity and duration) of the incident wavepacket:
\begin{eqnarray}\label{eq:it_product}
&&I \tau = \varkappa \frac{m_e c^2}{8} \frac{n_e \lambda_w}{\gamma_w R} \nonumber \\
&&= \varkappa \frac{3^{4/3} m_e c^2}{2 \Gamma^2\left( 2 / 3 \right)} \left(\frac{\omega_0}{\omega_{pe}} \right)^{8/3} \gamma_w^{1/3} n_e \lambda_w.
\end{eqnarray}
As can be seen from this formula, even a low intensity electromagnetic wavepacket is able to destroy the mirror, if it is sufficiently long. However, the relativistic mirror realized with the Langmuir wave consists of electrons that are continuously flowing through it. Consequently, the structure of the electron density spike is being refreshed every moment in time. Thus, the applicability of the model given by Eqs.~(\ref{eq:momentum_conservation}) - (\ref{eq:energies}) is better for a short-time interaction and sufficiently large electromagnetic wave intensity. We intrepret the threshold of Eq.~(\ref{eq:it_product}) as a condition for the minimum wavepacket duration required for a recoil effect,
\begin{eqnarray}\label{eq:tau_min}
\tau_{min} = \varkappa \frac{3^{4/3} m_e c^2}{2 \Gamma^2\left( 2 / 3 \right)} \left(\frac{\omega_0}{\omega_{pe}} \right)^{8/3} \gamma_w^{1/3} \frac{n_e \lambda_w}{I}.
\end{eqnarray}
In this interpretation, the incident wavepacket intensity becomes the main critical parameter for the recoil effects. Below, we investigate the applicability of the model and, in particular, Eq.~(\ref{eq:tau_min}) by PIC simulations.

\section{\label{sec:pic_simulations} Particle-in-cell simulations}

The properties of relativistic mirrors realized with strongly nonlinear Langmuir waves in underdense plasmas are studied numerically by means of PIC simulations in a 1D Cartesian geometry. The 1D configuration is sufficient for the investigation of the Langmuir wave interaction with a counter-propagating laser pulse and beneficial in view of the necessity of resolving frequency upshifted electromagnetic radiation according to Eq.~(\ref{eq:upshift_factor_plasma}). The results can be extrapolated to higher dimensions considering laser pulses with a wide focal spot. The simulations are performed using the fully relativistic electromagnetic PIC EPOCH code \cite{Arber2015}.

\subsection{\label{sec:pic_setup} Simulation Setup}

\begin{figure}
	\includegraphics[width=0.95\columnwidth]{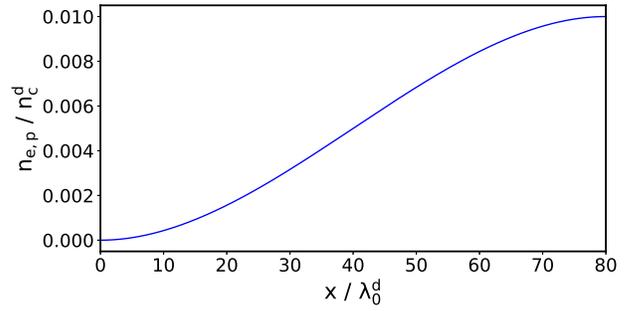}
	\caption{\label{fig:density_ramp} Plot of the electron and proton density ramp profile used in the simulations.}
\end{figure}

The laser pulse that drives the Langmuir wave (from here on referred to as the "driver") enters the simulation domain from the left boundary and propagates in the $ +x $ direction. The laser pulse that undergoes the reflection from the Langmuir wave (from here on referred to as the "source") enters from the right and propagates in the opposite (i.e. $ -x $) direction. In the following, we use the superscripts "$ d $", "$ s $" and "$ r $" to denote the quantities which characterize the driver, the source and the reflected pulse, respectively.

\begin{figure*}
	\includegraphics[width=0.9\linewidth]{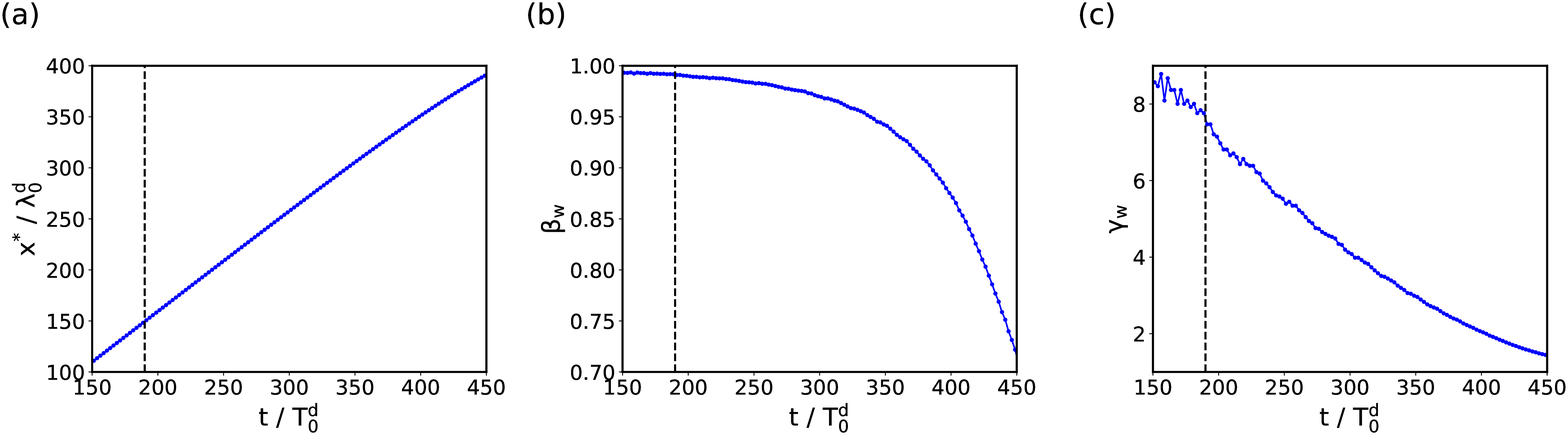}
	\caption{\label{fig:beta_gamma} The evolution of parameters of the first electron density spike of the Langmuir wave behind the driver in time. (a) The motion of the spike in the $ x - t $ plane, (b) the normalized phase velocity of the spike $ \beta_w $ and (c) the corresponding relativistic Lorentz factor $ \gamma_w $. The black dashed line marks the instant, when the Langmuir wave breaks.}
\end{figure*}

The driver is characterized by a wavelength in vacuum $ \lambda_{0}^d = 2 \pi c / \omega_{0}^d $, where $ \omega_{0}^d $ is its angular frequency, and by the normalized amplitude $ a_0^d = 10 $ defined as $ a_0^d = e E_0^d / (m_e \omega_0^d c) $, where $ E_0^d $ is amplitude of the electric field in vacuum. Its temporal profile is Gaussian with a full-width-at-half-maximum duration $ \tau_0^d = 10 \ T_0^d $, where $ T_0^d = \lambda_{0}^d / c $ is the driver pulse cycle period. The values of $ a_0^d $ and $ \tau_0^d $ are set so that they are optimal for the Langmuir wave generation \cite{Bulanov2016, Esarey2009}; the driver amplitude $ a_0^d $ is set to be sufficiently high in order to excite a large amplitude nonlinear wave which breaks in a controlled way and the driver duration $ \tau_0^d $ is chosen such that the wave is excited resonantly (i.e. $ c \tau_0^d \approx \lambda_w / 2 $). The driver is linearly polarized with the electric field directed along the $ y $-axis.

The wavelength of the source pulse is $ \lambda_0^s = 5 \ \lambda_0^d $. By this choice we keep $ \lambda_0^s $ sufficiently short so that the effects of plasma dispersion on the source are not significant, but long enough to substantially reduce the computational demands of the simulations. The source has a semi-infinite flat-top temporal profile which allows us to analyze the simulation results more clearly. The normalized amplitude of the source, $ a_0^s $, is varied in the simulations in order to thoroughly describe its impact on the reflection from the Langmuir wave. The source is linearly polarized in the direction perpendicular to the driver polarization (i.e. along the $ z $-axis), thus its electromagnetic field can be clearly distinguished.

Both laser pulses, the driver and the source, propagate in a pre-ionized uniform hydrogen plasma with electron and proton densities $ n_{e, p}^{0} = 10^{-2} \ n_c^d $, where $ n_c^d = m_e (\omega_0^d)^2 / (4 \pi e^2) $ is the critical plasma density with respect to the driver pulse. A smooth ramp is added to the left side of the target in order to reduce the effect of wave-breaking from a sharp rising plasma edge \cite{Bulanov1990}. The ramp is defined by the function $ n_{e, p} \left( x \right) = n_{e, p}^{0} \ (3 - 2 (x - x_1)/(x_2 - x_1)) ((x - x_1)/(x_2 - x_1))^2 $, where $ x \in \left[ x_1, \ x_2 \right] $. The values $ x_1 = 0 $ and $ x_2 = 80 \ \lambda_0^d $ have proven to provide a sufficiently smooth transition (one can see the plot of the density ramp $ n_{e, p} \left( x \right) $ in Fig.~\ref{fig:density_ramp}). The plasma is cold and collisionless. The electrons and protons are represented by quasi-particles with a triangular shape function. The number of quasi-particles per cell is $ 10 $ for both particle species.

The simulations utilize a moving window technique \cite{Fidel1997} which allows to substantially decrease the length of the simulation domain. For this, the EPOCH code was modified in order to continuously introduce source pulse at the right boundary of the moving widow. The length of the simulation window is $ 80 \ \lambda_0^d $ and it moves along the driver propagation direction at a velocity equal to $ c $. The resolution of the Cartesian grid is $ 30 $ cells per theoretically estimated wavelength of the reflected radiation $ \lambda^r $. The value of $ \lambda^r $ is calculated using Eq.~(\ref{eq:upshift_factor}), where we estimate $ \gamma_M \approx \omega_0^d / \omega_{pe} $. The simulation domain thus contains $ 1.92 \times 10^{5} $ cells in total and the simulation time is set to $ 450 \ T_0^d $. The electromagnetic fields are calculated using the standard second-order Yee solver \cite{Yee1966} with the CFL number \cite{Courant1928} equal to $ 0.99 $. Absorbing boundary conditions are applied on each of the simulation domain sides for both the electromagnetic field and particles.

\subsection{\label{sec:pic_results} Simulation Results}

First, we present the results of the simulation where the normalized amplitude of the source is relatively low, $ a_0^s = 10^{-4} $, in order to avoid recoil effects and significant distortions of the Langmuir wave. The driver pulse starts to excite the Langmuir wave as soon as it enters the plasma. When the driver reaches the uniform plasma density region, the Langmuir wave takes the form of sharp electron density spikes separated by cavities. We consider the properties of the first electron density spike of the Langmuir wave formed behind the driver, which serves as a relativistic mirror. The first important parameter of the density spike in our study is its phase velocity because it determines the magnitudes of the carrier frequency upshift and electric field amplification of the reflected wave (see Eqs.~(\ref{eq:upshift_factor}) and (\ref{eq:amplification_factor})). 

\begin{figure*}
	\includegraphics[width=0.9\linewidth]{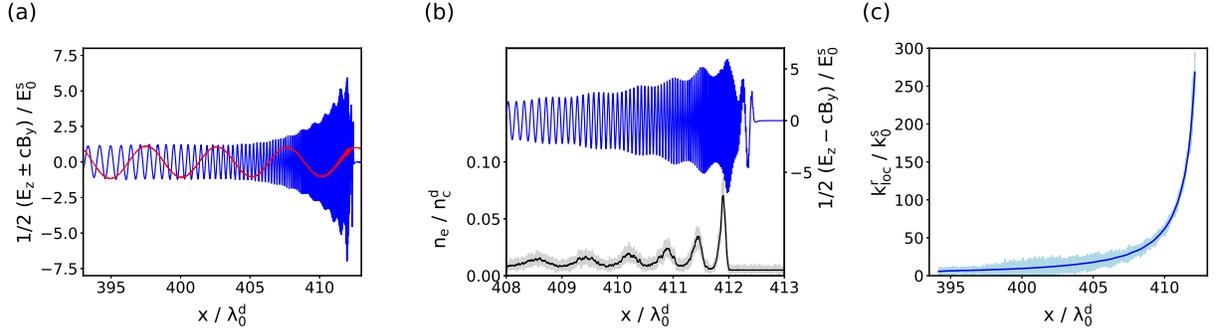}
	\caption{\label{fig:refl_wave} (a) Electromagnetic radiation incident at (red) and reflected from (blue) the first electron density spike of the Langmuir wave behind the driver, (b) detail of the reflected electromagnetic wave (blue) with modulations correlated with the electron density (black) and (c) the evolution of the local carrier wavenumber of the reflected electromagnetic wave showing a positive chirp due to the Langmuir wave deceleration. In (b) and (c), the simulation data (light gray and light blue) are smoothed using the Savitzky-Golay filter \cite{Savitzky1964} (black and blue).}
\end{figure*}

Fig.~\ref{fig:beta_gamma}(a) shows the evolution of the first electron density spike of the Langmuir wave behind the driver in the $ x - t $ plane, (b) its normalized phase velocity $ \beta_w $ and (c) the corresponding relativistic Lorentz factor $ \gamma_w $. The moment of wave-breaking, $ t \approx 190 \ T_0^d $, is in Fig.~\ref{fig:beta_gamma} denoted by black dashed lines. It corresponds to the limit given by Eq.~(\ref{eq:akhiezer-polovin}). At this moment, the electron density spike is centered around the point $ x \approx 150 \ \lambda_0^d $. The reflectivity of the Langmuir wave becomes significant when the wave is closer to breaking \cite{Bulanov2013}. After the wave-breaking, it is determined not only by the properties of the regular Langmuir wave, but also by the properties of the injected electrons. From Fig.~\ref{fig:beta_gamma}, it can be clearly seen that the Langmuir wave decelerates in uniform plasma, which is partially caused due to nonlinear energy depletion of the driver \cite{Bulanov1992} and due to the wave-breaking.

The reflected electromagnetic radiation is shown in Fig.~\ref{fig:refl_wave}(a). As can be seen, its amplitude is modulated. The modulations are caused by the electrons injected into the accelerating phase of the wakefield after the wave-breaking, which is shown in Fig.~\ref{fig:refl_wave}(b). Fig.~\ref{fig:refl_wave}(c) displays the local carrier wavenumber of the reflected pulse. To obtain the local carrier wavenumber at any point in a reflected wavepacket, we first find the analytic signal from the original signal using the Hilbert transform \cite{King2009}. The local carrier wavenumber is then obtained by differentiating the local phase (which corresponds to the phase angle of the analytic signal) with respect to $ x $. It can be clearly seen that the reflected signal has a positive chirp which corresponds to the mirror deceleration. The wavelength of the reflected signal $ \lambda^r $ ranges from $ \approx 0.17 \ \lambda_0^s $ to $ \approx 3.3 \times 10^{-3} \ \lambda_0^s $, hence the upshift factor with respect to $ \omega_0^s $ ranges from $ 6 $ to $ 298 $. Due to the effects of plasma dispersion, however, the wavelength of the source pulse interacting with the electron density spike is slightly larger than the vacuum wavelength, $ \lambda^s \approx 1.04 \ \lambda_0^s $. Thus the maximum factor of the frequency upshift with respect to $ \omega^s $ is about $ 310 $. From this frequency upshift factor using Eq.~(\ref{eq:upshift_factor_plasma}) we can estimate the relativistic Lorentz factor of the electron density spike as $ \gamma_w \approx 9.2 $, which corresponds to the instant of time $ t \approx 140 \ T_0^d $. 

\begin{figure*}
	\includegraphics[width=0.9\linewidth]{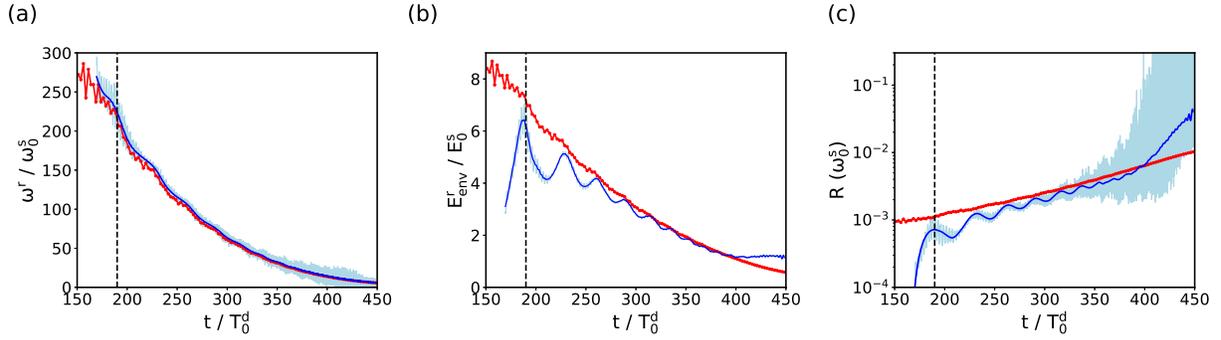}
	\caption{\label{fig:upshift_amplification} The properties of the reflection from the first electron density spike of the Langmuir wave behind the driver. (a) The frequency upshift factor, (b) the electric field amplification factor and (c) the instantaneous reflection coefficient in terms of photon number. The simulation data (light blue) are smoothed using the Savitzky-Golay filter \cite{Savitzky1964} (blue) and compared to analytical estimates (red). The black dashed line marks the instant, when the Langmuir wave breaks.}
\end{figure*}

Using the dependence of the local carrier wavenumber of the reflected pulse on the electron density spike coordinate $ k_{loc}^r (x) $ and the dependence of the spike coordinate on time $ x^{*} (t) $, we obtain the time dependence of the frequency upshift factor of the reflected pulse,
\begin{equation*}
\frac{\omega^r}{\omega_0^s} = \frac{k_0^s}{k_{loc}^r(x^{*}(t))}
\end{equation*}
As seen in Fig.~\ref{fig:upshift_amplification}(a), the frequency upshift factor obtained in this way very well agrees with the calculation using Eq.~(\ref{eq:upshift_factor_plasma}) and the relativistic Lorentz factor of the electron density spike $ \gamma_w $ shown in Fig.~\ref{fig:beta_gamma}(c).

Using the dependence of the reflected pulse electric field envelope amplitude on the spike coordinate $ E_{env}^r(x^{*}(t)) $, we obtain the time dependence of the electric field amplification factor,
\begin{equation*}
\frac{E^r}{E_0^s} = \frac{E_{env}^r(x^{*}(t))}{E_0^s}.
\end{equation*}
As seen in Fig.~\ref{fig:upshift_amplification}(b), the electric field amplification factor obtained in this way shows again fairly good conformity with the calculation using Eqs.~(\ref{eq:upshift_factor_plasma}) and (\ref{eq:refl_coef}) and the relativistic Lorentz factor of the electron density spike $ \gamma_w $ shown in Fig.~\ref{fig:beta_gamma}(c). We find that the electric field amplification factor reaches its maximum at the moment of wave-breaking, with the electric field of the reflected pulse being amplified more than 6 times.

Using the factors of the frequency upshift and the electric field amplification of the reflected pulse shown in Fig.~\ref{fig:upshift_amplification}(a) and (b), we reconstruct the instantaneous reflection coefficient of the electron density spike in time, Fig.~\ref{fig:upshift_amplification}(c). For comparison, in Fig.~\ref{fig:upshift_amplification}(c) we also plot the instantaneous reflection coefficient computed using Eq.~(\ref{eq:refl_coef}) and the relativistic Lorentz factor $ \gamma_w $ shown in Fig.~\ref{fig:beta_gamma}(c). We find that the reflection coefficient in terms of photon number grows from $ \approx 10^{-3} $ at the moment of wave-breaking up to $ \approx 5 \times 10^{-2} $ at the end of the interaction.

\begin{figure}
	\includegraphics[width=0.95\columnwidth]{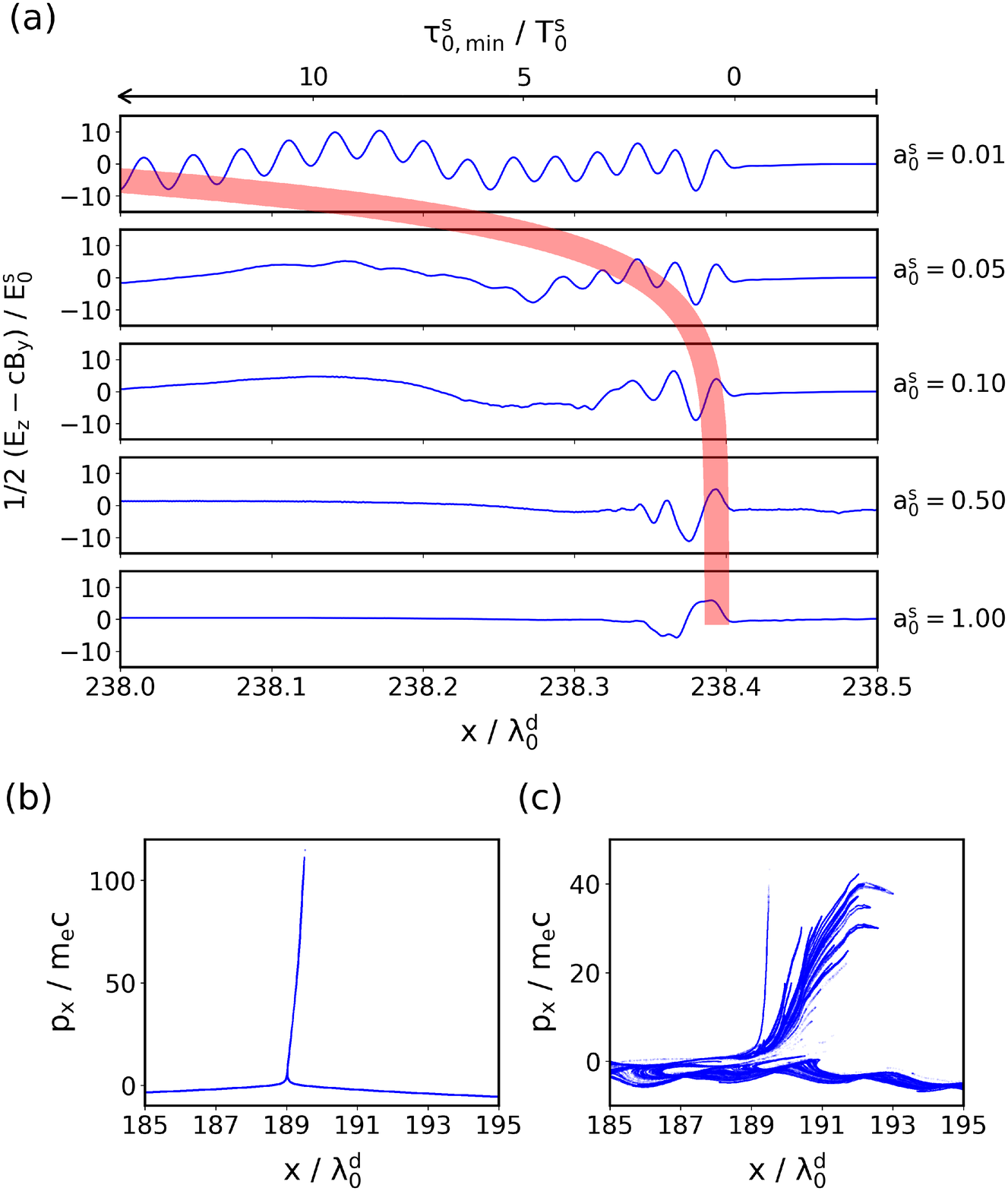}
	\caption{\label{fig:energy_threshold} Dependence of the reflected radiation on the amplitude of the source pulse. (a) Spatial profiles of the reflected wave for different source amplitudes. The thick red curve across the frames is the minimum source duration required for the recoil effect, $ \tau_{0, min}^s $, according to Eq.~(\ref{eq:tau_min}) for $ \varkappa = 1.5 \times 10^{-4} $. Phase space of electrons at $ t = 230 \ T_0^d $ for (b) $ a_0^s = 0.01 $ and (c) $ a_0^s = 1.0 $.}
\end{figure}

In order to investigate the recoil effects and explore the regimes around the threshold given by Eq.~(\ref{eq:it_product}), we increase the amplitude of the source. Now, the source pulse encounters the electron density spike at the moment of wave-breaking ($ t = 190 \ T_0^d $). Its normalized amplitude $ a_0^s $ is varied from $ 0.01 $ to $ 1 $. The reflected radiation for the simulated cases can be seen in Fig. \ref{fig:energy_threshold}(a). In the case of $ a_0^s = 0.01 $, the interaction corresponds to the weak incident pulse approximation and the impact of the source pulse is compensated by the electron flow that refreshes the structure of the density spike. For much larger amplitude, $ a_0^s = 1 $, only one cycle of the reflected wave is formed before the relativistic mirror is destroyed. Moreover, the radiation pressure of the source pulse in this case pushes the mirror back which results in lower factors of the frequency upshift and the electric field amplification.

The threshold~(\ref{eq:tau_min}) gives the minimal duration of the incident electromagnetic wave necessary to cause a significant recoil on the relativistic mirror. In terms of normalized quantities, this duration can be rewritten as
\begin{equation}\label{eq:tau_min_normalized}
\tau_{0, min}^s = \frac{\varkappa}{4} \frac{n_e \lambda_w \lambda_0^s}{\gamma_w R \left(a_0^s\right)^2},
\end{equation}
where $ \tau_{0, min}^s $ is normalized by $ T_0^s $, $ \lambda_0^s $ and $ \lambda_w $ by $ \lambda_0^d $ and $ n_e $ by $ n_c^d $. This quantity is shown in Fig.~\ref{fig:energy_threshold}(a), for different source pulse amplitudes and $ \varkappa = 1.5 \times 10^{-4} $. The value of the coefficient $ \varkappa $ is obtained from the comparison of the spatial profiles of the reflected wave for different incident wavepacket amplitudes. We assume that the duration $ \tau_{0,min}^s $ roughly corresponds to the time period where the reflected wave coincides with the weak-source approximation. We see that for $ a_0^s = 0.01 $ the reflected wave corresponds to the weak-source approximation and classical double Doppler effect. Here $ \tau_{0, min}^s $ is very large. For $ a_0^s = 1 $, the recoil effects are well pronounced; the spatial profile of the reflected wave deviates from the weak-source approximation almost immediately. In this case $ \tau_{0, min}^s $ is almost zero. Between $ a_0^s = 0.01 $ and $ a_0^s = 1 $, the properties of the spatial profile of the reflected wave correlate with the minimum source duration causing recoil effects given by Eq.~(\ref{eq:tau_min_normalized}), derived from the model (\ref{eq:momentum_conservation}) - (\ref{eq:energies}).

A time span needed for a density spike to be fully refreshed by the electron flow can be roughly estimated as $ t_{ref} \approx \lambda_w / v_g^d \approx 20.53 \ T_0^d $, where $ v_g^d $ is the group velocity of the driver pulse. During this time span, the density spike interacts approximately with $ 7.62 $ cycles of the source pulse. Therefore, if $ \tau_{0, min}^s > 7.62 \ T_0^s $ the impact of the source pulse on the density spike is compensated by the flow of electrons and the interaction corresponds to the weak-source approximation. Using Eq.~(\ref{eq:tau_min_normalized}) with $ \varkappa = 1.5 \times 10^{-4} $, this condition is equivalent to $ a_0^s < 0.026 $.

In Fig.~\ref{fig:energy_threshold}(b) and (c), one can see the phase space of electrons located in the density spike illustrating the importance of the recoil effects of the relativistic mirror for two different amplitudes of the source pulse. For relatively small amplitudes, the structure of the electron density spike and the injected electrons (appearing after wave-breaking) are not affected, Fig.~\ref{fig:energy_threshold}(b). When the intensity of the source pulse becomes sufficient to alter the motion of the electrons in the density spike, the spike splits into several layers, Fig.~\ref{fig:energy_threshold}(c). The disappearance of the periodic structure of the reflected electromagnetic wave seen in Fig.~\ref{fig:energy_threshold}(a) for $ a_0^s \geq 0.05 $ is partially due to destructive interference of waves reflected from the multi-layered structure of the split electron density spike and due to the recoil effects.

\section{\label{sec:conclusion} Conclusion}

We study recoil effects of relativistic mirrors in the form of strongly nonlinear Langmuir waves driven by short intense laser pulses in underdense plasmas. This is important for the question of the feasibility of relativistic mirrors for the development of compact and tunable sources of coherent short-wavelength radiation. Using analytical calculations and PIC simulations, we investigate the properties of the Langmuir wave as well as the reflected pulse. We also find the threshold for the energy of the laser pulse incident on the electron density spike above which the relativistic mirror undergoes significant recoil. 

We show that the Langmuir wave driven by a short intense laser pulse in uniform plasma decelerates and, therefore, the reflected radiation has a positive chirp. We find that the electric field amplification factor of the reflected radiation reaches its maximum at the moment of wave-breaking. In addition, our results show that for a given intensity of the source pulse there exists an optimal duration of the source pulse; longer-than-optimal pulses have lower reflected-to-incident energy ratio. Moreover, for a given Langmuir wave excited by the driver pulse there exists an optimal intensity of the source pulse which provides the most intense reflected wave with almost the same frequency upshift factor as in the weak-source approximation.

The sources of coherent high-brightness radiation with wavelengths ranging from x-ray to gamma-ray are of great demand for many practical applications in diverse fields. Relativistic mirrors in laser plasmas can give a promising alternative for the development of radiation sources with tunable parameters at significantly reduced size and cost, in comparison with conventional devices.


\begin{acknowledgments}
This work was supported by the project High Field Initiative (CZ.02.1.01/0.0/0.0/15\_003/0000449) from the European Regional Development Fund and by the project "IT4Innovations National Supercomputing Center – LM2015070" from The Ministry of Education, Youth and Sports of the Czech Republic. This work was in part funded by the UK EPSRC grants EP/G054950/1, EP/G056803/1, EP/G055165/1 and EP/M022463/1.
\end{acknowledgments}

\bibliography{refs}

\begin{thebibliography}{46}%
\makeatletter
\providecommand \@ifxundefined [1]{%
 \@ifx{#1\undefined}
}%
\providecommand \@ifnum [1]{%
 \ifnum #1\expandafter \@firstoftwo
 \else \expandafter \@secondoftwo
 \fi
}%
\providecommand \@ifx [1]{%
 \ifx #1\expandafter \@firstoftwo
 \else \expandafter \@secondoftwo
 \fi
}%
\providecommand \natexlab [1]{#1}%
\providecommand \enquote  [1]{``#1''}%
\providecommand \bibnamefont  [1]{#1}%
\providecommand \bibfnamefont [1]{#1}%
\providecommand \citenamefont [1]{#1}%
\providecommand \href@noop [0]{\@secondoftwo}%
\providecommand \href [0]{\begingroup \@sanitize@url \@href}%
\providecommand \@href[1]{\@@startlink{#1}\@@href}%
\providecommand \@@href[1]{\endgroup#1\@@endlink}%
\providecommand \@sanitize@url [0]{\catcode `\\12\catcode `\$12\catcode
  `\&12\catcode `\#12\catcode `\^12\catcode `\_12\catcode `\%12\relax}%
\providecommand \@@startlink[1]{}%
\providecommand \@@endlink[0]{}%
\providecommand \url  [0]{\begingroup\@sanitize@url \@url }%
\providecommand \@url [1]{\endgroup\@href {#1}{\urlprefix }}%
\providecommand \urlprefix  [0]{URL }%
\providecommand \Eprint [0]{\href }%
\providecommand \doibase [0]{http://dx.doi.org/}%
\providecommand \selectlanguage [0]{\@gobble}%
\providecommand \bibinfo  [0]{\@secondoftwo}%
\providecommand \bibfield  [0]{\@secondoftwo}%
\providecommand \translation [1]{[#1]}%
\providecommand \BibitemOpen [0]{}%
\providecommand \bibitemStop [0]{}%
\providecommand \bibitemNoStop [0]{.\EOS\space}%
\providecommand \EOS [0]{\spacefactor3000\relax}%
\providecommand \BibitemShut  [1]{\csname bibitem#1\endcsname}%
\let\auto@bib@innerbib\@empty
\bibitem [{\citenamefont {Einstein}(1905)}]{Einstein1905}%
  \BibitemOpen
  \bibfield  {author} {\bibinfo {author} {\bibfnamefont {A.}~\bibnamefont
  {Einstein}},\ }\href {\doibase 10.1002/andp.19053221004} {\bibfield
  {journal} {\bibinfo  {journal} {Annalen der Physik}\ }\textbf {\bibinfo
  {volume} {322}},\ \bibinfo {pages} {891} (\bibinfo {year}
  {1905})}\BibitemShut {NoStop}%
\bibitem [{\citenamefont {Bulanov}\ \emph {et~al.}(2013)\citenamefont
  {Bulanov}, \citenamefont {Esirkepov}, \citenamefont {Kando}, \citenamefont
  {Pirozhkov},\ and\ \citenamefont {Rosanov}}]{Bulanov2013}%
  \BibitemOpen
  \bibfield  {author} {\bibinfo {author} {\bibfnamefont {S.~V.}\ \bibnamefont
  {Bulanov}}, \bibinfo {author} {\bibfnamefont {T.~Z.}\ \bibnamefont
  {Esirkepov}}, \bibinfo {author} {\bibfnamefont {M.}~\bibnamefont {Kando}},
  \bibinfo {author} {\bibfnamefont {A.~S.}\ \bibnamefont {Pirozhkov}}, \ and\
  \bibinfo {author} {\bibfnamefont {N.~N.}\ \bibnamefont {Rosanov}},\ }\href
  {\doibase 10.3367/ufne.0183.201305a.0449} {\bibfield  {journal} {\bibinfo
  {journal} {Physics-Uspekhi}\ }\textbf {\bibinfo {volume} {56}},\ \bibinfo
  {pages} {429} (\bibinfo {year} {2013})}\BibitemShut {NoStop}%
\bibitem [{\citenamefont {Kulagin}\ \emph
  {et~al.}(2007{\natexlab{a}})\citenamefont {Kulagin}, \citenamefont
  {Cherepenin}, \citenamefont {Hur},\ and\ \citenamefont {Suk}}]{Kulagin2007}%
  \BibitemOpen
  \bibfield  {author} {\bibinfo {author} {\bibfnamefont {V.~V.}\ \bibnamefont
  {Kulagin}}, \bibinfo {author} {\bibfnamefont {V.~A.}\ \bibnamefont
  {Cherepenin}}, \bibinfo {author} {\bibfnamefont {M.~S.}\ \bibnamefont {Hur}},
  \ and\ \bibinfo {author} {\bibfnamefont {H.}~\bibnamefont {Suk}},\ }\href
  {\doibase 10.1063/1.2799164} {\bibfield  {journal} {\bibinfo  {journal}
  {Physics of Plasmas}\ }\textbf {\bibinfo {volume} {14}},\ \bibinfo {pages}
  {113101} (\bibinfo {year} {2007}{\natexlab{a}})}\BibitemShut {NoStop}%
\bibitem [{\citenamefont {Kulagin}\ \emph
  {et~al.}(2007{\natexlab{b}})\citenamefont {Kulagin}, \citenamefont
  {Cherepenin}, \citenamefont {Hur},\ and\ \citenamefont {Suk}}]{Kulagin2007a}%
  \BibitemOpen
  \bibfield  {author} {\bibinfo {author} {\bibfnamefont {V.~V.}\ \bibnamefont
  {Kulagin}}, \bibinfo {author} {\bibfnamefont {V.~A.}\ \bibnamefont
  {Cherepenin}}, \bibinfo {author} {\bibfnamefont {M.~S.}\ \bibnamefont {Hur}},
  \ and\ \bibinfo {author} {\bibfnamefont {H.}~\bibnamefont {Suk}},\ }\href
  {\doibase 10.1103/PhysRevLett.99.124801} {\bibfield  {journal} {\bibinfo
  {journal} {Physical Review Letters}\ }\textbf {\bibinfo {volume} {99}},\
  \bibinfo {pages} {124801} (\bibinfo {year} {2007}{\natexlab{b}})}\BibitemShut
  {NoStop}%
\bibitem [{\citenamefont {Esirkepov}\ \emph {et~al.}(2009)\citenamefont
  {Esirkepov}, \citenamefont {Bulanov}, \citenamefont {Kando}, \citenamefont
  {Pirozhkov},\ and\ \citenamefont {Zhidkov}}]{Esirkepov2009}%
  \BibitemOpen
  \bibfield  {author} {\bibinfo {author} {\bibfnamefont {T.~Z.}\ \bibnamefont
  {Esirkepov}}, \bibinfo {author} {\bibfnamefont {S.~V.}\ \bibnamefont
  {Bulanov}}, \bibinfo {author} {\bibfnamefont {M.}~\bibnamefont {Kando}},
  \bibinfo {author} {\bibfnamefont {A.~S.}\ \bibnamefont {Pirozhkov}}, \ and\
  \bibinfo {author} {\bibfnamefont {A.~G.}\ \bibnamefont {Zhidkov}},\ }\href
  {\doibase 10.1103/PhysRevLett.103.025002} {\bibfield  {journal} {\bibinfo
  {journal} {Physical Review Letters}\ }\textbf {\bibinfo {volume} {103}},\
  \bibinfo {pages} {25002} (\bibinfo {year} {2009})}\BibitemShut {NoStop}%
\bibitem [{\citenamefont {Meyer-Ter-Vehn}\ and\ \citenamefont
  {Wu}(2009)}]{Meyer-ter-Vehn2009}%
  \BibitemOpen
  \bibfield  {author} {\bibinfo {author} {\bibfnamefont {J.}~\bibnamefont
  {Meyer-Ter-Vehn}}\ and\ \bibinfo {author} {\bibfnamefont {H.~C.}\
  \bibnamefont {Wu}},\ }\href {\doibase 10.1140/epjd/e2009-00081-1} {\bibfield
  {journal} {\bibinfo  {journal} {European Physical Journal D}\ }\textbf
  {\bibinfo {volume} {55}},\ \bibinfo {pages} {433} (\bibinfo {year} {2009})},\
  \Eprint {http://arxiv.org/abs/0812.0710} {arXiv:0812.0710} \BibitemShut
  {NoStop}%
\bibitem [{\citenamefont {Bulanov}\ \emph {et~al.}(2010)\citenamefont
  {Bulanov}, \citenamefont {Maksimchuk}, \citenamefont {Krushelnick},
  \citenamefont {Popov}, \citenamefont {Bychenkov},\ and\ \citenamefont
  {Rozmus}}]{Bulanov2010}%
  \BibitemOpen
  \bibfield  {author} {\bibinfo {author} {\bibfnamefont {S.~S.}\ \bibnamefont
  {Bulanov}}, \bibinfo {author} {\bibfnamefont {A.}~\bibnamefont {Maksimchuk}},
  \bibinfo {author} {\bibfnamefont {K.}~\bibnamefont {Krushelnick}}, \bibinfo
  {author} {\bibfnamefont {K.~I.}\ \bibnamefont {Popov}}, \bibinfo {author}
  {\bibfnamefont {V.~Y.}\ \bibnamefont {Bychenkov}}, \ and\ \bibinfo {author}
  {\bibfnamefont {W.}~\bibnamefont {Rozmus}},\ }\href {\doibase
  10.1016/j.physleta.2009.11.009} {\bibfield  {journal} {\bibinfo  {journal}
  {Physics Letters, Section A: General, Atomic and Solid State Physics}\
  }\textbf {\bibinfo {volume} {374}},\ \bibinfo {pages} {476} (\bibinfo {year}
  {2010})},\ \Eprint {http://arxiv.org/abs/0911.4767} {arXiv:0911.4767}
  \BibitemShut {NoStop}%
\bibitem [{\citenamefont {Wu}\ \emph {et~al.}(2010)\citenamefont {Wu},
  \citenamefont {Meyer-Ter-Vehn}, \citenamefont {Fern{\'{a}}ndez},\ and\
  \citenamefont {Hegelich}}]{Wu2010}%
  \BibitemOpen
  \bibfield  {author} {\bibinfo {author} {\bibfnamefont {H.~C.}\ \bibnamefont
  {Wu}}, \bibinfo {author} {\bibfnamefont {J.}~\bibnamefont {Meyer-Ter-Vehn}},
  \bibinfo {author} {\bibfnamefont {J.}~\bibnamefont {Fern{\'{a}}ndez}}, \ and\
  \bibinfo {author} {\bibfnamefont {B.~M.}\ \bibnamefont {Hegelich}},\ }\href
  {\doibase 10.1103/PhysRevLett.104.234801} {\bibfield  {journal} {\bibinfo
  {journal} {Physical Review Letters}\ }\textbf {\bibinfo {volume} {104}},\
  \bibinfo {pages} {234801} (\bibinfo {year} {2010})}\BibitemShut {NoStop}%
\bibitem [{\citenamefont {Wu}\ \emph {et~al.}(2011)\citenamefont {Wu},
  \citenamefont {Meyer-Ter-Vehn}, \citenamefont {Hegelich},\ and\ \citenamefont
  {Fern{\'{a}}ndez}}]{Wu2011}%
  \BibitemOpen
  \bibfield  {author} {\bibinfo {author} {\bibfnamefont {H.~C.}\ \bibnamefont
  {Wu}}, \bibinfo {author} {\bibfnamefont {J.}~\bibnamefont {Meyer-Ter-Vehn}},
  \bibinfo {author} {\bibfnamefont {B.~M.}\ \bibnamefont {Hegelich}}, \ and\
  \bibinfo {author} {\bibfnamefont {J.~C.}\ \bibnamefont {Fern{\'{a}}ndez}},\
  }\href {\doibase 10.1103/PhysRevSTAB.14.070702} {\bibfield  {journal}
  {\bibinfo  {journal} {Physical Review Special Topics - Accelerators and
  Beams}\ }\textbf {\bibinfo {volume} {14}},\ \bibinfo {pages} {70702}
  (\bibinfo {year} {2011})}\BibitemShut {NoStop}%
\bibitem [{\citenamefont {Wu}\ and\ \citenamefont {Meyer-ter
  Vehn}(2012)}]{Wu2012}%
  \BibitemOpen
  \bibfield  {author} {\bibinfo {author} {\bibfnamefont {H.-C.}\ \bibnamefont
  {Wu}}\ and\ \bibinfo {author} {\bibfnamefont {J.}~\bibnamefont {Meyer-ter
  Vehn}},\ }\href {https://doi.org/10.1038/nphoton.2012.76} {\bibfield
  {journal} {\bibinfo  {journal} {Nature Photonics}\ }\textbf {\bibinfo
  {volume} {6}},\ \bibinfo {pages} {304} (\bibinfo {year} {2012})}\BibitemShut
  {NoStop}%
\bibitem [{\citenamefont {Andreev}, \citenamefont {Platonov},\ and\
  \citenamefont {Sadykova}(2013)}]{Andreev2013}%
  \BibitemOpen
  \bibfield  {author} {\bibinfo {author} {\bibfnamefont {A.}~\bibnamefont
  {Andreev}}, \bibinfo {author} {\bibfnamefont {K.}~\bibnamefont {Platonov}}, \
  and\ \bibinfo {author} {\bibfnamefont {S.}~\bibnamefont {Sadykova}},\ }\href
  {\doibase 10.3390/app3010094} {\bibfield  {journal} {\bibinfo  {journal}
  {Applied Sciences}\ }\textbf {\bibinfo {volume} {3}},\ \bibinfo {pages} {94}
  (\bibinfo {year} {2013})}\BibitemShut {NoStop}%
\bibitem [{\citenamefont {Kiefer}\ \emph {et~al.}(2013)\citenamefont {Kiefer},
  \citenamefont {Yeung}, \citenamefont {Dzelzainis}, \citenamefont {Foster},
  \citenamefont {Rykovanov}, \citenamefont {Lewis}, \citenamefont
  {Marjoribanks}, \citenamefont {Ruhl}, \citenamefont {Habs}, \citenamefont
  {Schreiber}, \citenamefont {Zepf},\ and\ \citenamefont
  {Dromey}}]{Kiefer2013}%
  \BibitemOpen
  \bibfield  {author} {\bibinfo {author} {\bibfnamefont {D.}~\bibnamefont
  {Kiefer}}, \bibinfo {author} {\bibfnamefont {M.}~\bibnamefont {Yeung}},
  \bibinfo {author} {\bibfnamefont {T.}~\bibnamefont {Dzelzainis}}, \bibinfo
  {author} {\bibfnamefont {P.~S.}\ \bibnamefont {Foster}}, \bibinfo {author}
  {\bibfnamefont {S.~G.}\ \bibnamefont {Rykovanov}}, \bibinfo {author}
  {\bibfnamefont {C.~L.}\ \bibnamefont {Lewis}}, \bibinfo {author}
  {\bibfnamefont {R.~S.}\ \bibnamefont {Marjoribanks}}, \bibinfo {author}
  {\bibfnamefont {H.}~\bibnamefont {Ruhl}}, \bibinfo {author} {\bibfnamefont
  {D.}~\bibnamefont {Habs}}, \bibinfo {author} {\bibfnamefont {J.}~\bibnamefont
  {Schreiber}}, \bibinfo {author} {\bibfnamefont {M.}~\bibnamefont {Zepf}}, \
  and\ \bibinfo {author} {\bibfnamefont {B.}~\bibnamefont {Dromey}},\ }\href
  {\doibase 10.1038/ncomms2775} {\bibfield  {journal} {\bibinfo  {journal}
  {Nature Communications}\ }\textbf {\bibinfo {volume} {4}},\ \bibinfo {pages}
  {1763} (\bibinfo {year} {2013})}\BibitemShut {NoStop}%
\bibitem [{\citenamefont {Ma}\ \emph {et~al.}(2014)\citenamefont {Ma},
  \citenamefont {Bin}, \citenamefont {Wang}, \citenamefont {Yeung},
  \citenamefont {Kreuzer}, \citenamefont {Streeter}, \citenamefont {Foster},
  \citenamefont {Cousens}, \citenamefont {Kiefer}, \citenamefont {Dromey},
  \citenamefont {Yan}, \citenamefont {Meyer-Ter-Vehn}, \citenamefont {Zepf},\
  and\ \citenamefont {Schreiber}}]{Ma2014}%
  \BibitemOpen
  \bibfield  {author} {\bibinfo {author} {\bibfnamefont {W.~J.}\ \bibnamefont
  {Ma}}, \bibinfo {author} {\bibfnamefont {J.~H.}\ \bibnamefont {Bin}},
  \bibinfo {author} {\bibfnamefont {H.~Y.}\ \bibnamefont {Wang}}, \bibinfo
  {author} {\bibfnamefont {M.}~\bibnamefont {Yeung}}, \bibinfo {author}
  {\bibfnamefont {C.}~\bibnamefont {Kreuzer}}, \bibinfo {author} {\bibfnamefont
  {M.}~\bibnamefont {Streeter}}, \bibinfo {author} {\bibfnamefont {P.~S.}\
  \bibnamefont {Foster}}, \bibinfo {author} {\bibfnamefont {S.}~\bibnamefont
  {Cousens}}, \bibinfo {author} {\bibfnamefont {D.}~\bibnamefont {Kiefer}},
  \bibinfo {author} {\bibfnamefont {B.}~\bibnamefont {Dromey}}, \bibinfo
  {author} {\bibfnamefont {X.~Q.}\ \bibnamefont {Yan}}, \bibinfo {author}
  {\bibfnamefont {J.}~\bibnamefont {Meyer-Ter-Vehn}}, \bibinfo {author}
  {\bibfnamefont {M.}~\bibnamefont {Zepf}}, \ and\ \bibinfo {author}
  {\bibfnamefont {J.}~\bibnamefont {Schreiber}},\ }\href {\doibase
  10.1103/PhysRevLett.113.235002} {\bibfield  {journal} {\bibinfo  {journal}
  {Physical Review Letters}\ }\textbf {\bibinfo {volume} {113}},\ \bibinfo
  {pages} {235002} (\bibinfo {year} {2014})}\BibitemShut {NoStop}%
\bibitem [{\citenamefont {Bulanov}, \citenamefont {Naumova},\ and\
  \citenamefont {Pegoraro}(1994)}]{Bulanov1994}%
  \BibitemOpen
  \bibfield  {author} {\bibinfo {author} {\bibfnamefont {S.~V.}\ \bibnamefont
  {Bulanov}}, \bibinfo {author} {\bibfnamefont {N.~M.}\ \bibnamefont
  {Naumova}}, \ and\ \bibinfo {author} {\bibfnamefont {F.}~\bibnamefont
  {Pegoraro}},\ }\href {\doibase 10.1063/1.870766} {\bibfield  {journal}
  {\bibinfo  {journal} {Physics of Plasmas}\ }\textbf {\bibinfo {volume} {1}},\
  \bibinfo {pages} {745} (\bibinfo {year} {1994})}\BibitemShut {NoStop}%
\bibitem [{\citenamefont {Lichters}, \citenamefont {Meyer-ter Vehn},\ and\
  \citenamefont {Pukhov}(1996)}]{Lichters1996}%
  \BibitemOpen
  \bibfield  {author} {\bibinfo {author} {\bibfnamefont {R.}~\bibnamefont
  {Lichters}}, \bibinfo {author} {\bibfnamefont {J.}~\bibnamefont {Meyer-ter
  Vehn}}, \ and\ \bibinfo {author} {\bibfnamefont {A.}~\bibnamefont {Pukhov}},\
  }\href {\doibase 10.1063/1.871619} {\bibfield  {journal} {\bibinfo  {journal}
  {Physics of Plasmas}\ }\textbf {\bibinfo {volume} {3}},\ \bibinfo {pages}
  {3425} (\bibinfo {year} {1996})}\BibitemShut {NoStop}%
\bibitem [{\citenamefont {Naumova}\ \emph {et~al.}(2004)\citenamefont
  {Naumova}, \citenamefont {Nees}, \citenamefont {Sokolov}, \citenamefont
  {Hou},\ and\ \citenamefont {Mourou}}]{Naumova2004}%
  \BibitemOpen
  \bibfield  {author} {\bibinfo {author} {\bibfnamefont {N.~M.}\ \bibnamefont
  {Naumova}}, \bibinfo {author} {\bibfnamefont {J.~A.}\ \bibnamefont {Nees}},
  \bibinfo {author} {\bibfnamefont {I.~V.}\ \bibnamefont {Sokolov}}, \bibinfo
  {author} {\bibfnamefont {B.}~\bibnamefont {Hou}}, \ and\ \bibinfo {author}
  {\bibfnamefont {G.~A.}\ \bibnamefont {Mourou}},\ }\href {\doibase
  10.1103/PhysRevLett.92.063902} {\bibfield  {journal} {\bibinfo  {journal}
  {Physical Review Letters}\ }\textbf {\bibinfo {volume} {92}},\ \bibinfo
  {pages} {63902} (\bibinfo {year} {2004})}\BibitemShut {NoStop}%
\bibitem [{\citenamefont {Baeva}, \citenamefont {Gordienko},\ and\
  \citenamefont {Pukhov}(2006)}]{Baeva2006}%
  \BibitemOpen
  \bibfield  {author} {\bibinfo {author} {\bibfnamefont {T.}~\bibnamefont
  {Baeva}}, \bibinfo {author} {\bibfnamefont {S.}~\bibnamefont {Gordienko}}, \
  and\ \bibinfo {author} {\bibfnamefont {A.}~\bibnamefont {Pukhov}},\ }\href
  {\doibase 10.1103/PhysRevE.74.046404} {\bibfield  {journal} {\bibinfo
  {journal} {Physical Review E - Statistical, Nonlinear, and Soft Matter
  Physics}\ }\textbf {\bibinfo {volume} {74}},\ \bibinfo {pages} {46404}
  (\bibinfo {year} {2006})},\ \Eprint {http://arxiv.org/abs/0604228}
  {arXiv:0604228 [physics]} \BibitemShut {NoStop}%
\bibitem [{\citenamefont {Wheeler}\ \emph {et~al.}(2012)\citenamefont
  {Wheeler}, \citenamefont {Borot}, \citenamefont {Malvache}, \citenamefont
  {Ricci}, \citenamefont {Jullien}, \citenamefont {Lopez-Martens},
  \citenamefont {Monchoc{\'{e}}}, \citenamefont {Vincenti},\ and\ \citenamefont
  {Qu{\'{e}}r{\'{e}}}}]{Wheeler2012}%
  \BibitemOpen
  \bibfield  {author} {\bibinfo {author} {\bibfnamefont {J.}~\bibnamefont
  {Wheeler}}, \bibinfo {author} {\bibfnamefont {A.}~\bibnamefont {Borot}},
  \bibinfo {author} {\bibfnamefont {A.}~\bibnamefont {Malvache}}, \bibinfo
  {author} {\bibfnamefont {A.}~\bibnamefont {Ricci}}, \bibinfo {author}
  {\bibfnamefont {A.}~\bibnamefont {Jullien}}, \bibinfo {author} {\bibfnamefont
  {R.}~\bibnamefont {Lopez-Martens}}, \bibinfo {author} {\bibfnamefont
  {S.}~\bibnamefont {Monchoc{\'{e}}}}, \bibinfo {author} {\bibfnamefont
  {H.}~\bibnamefont {Vincenti}}, \ and\ \bibinfo {author} {\bibfnamefont
  {F.}~\bibnamefont {Qu{\'{e}}r{\'{e}}}},\ }\href {\doibase
  10.1364/qels.2012.qth5b.9} {\bibfield  {journal} {\bibinfo  {journal} {Optics
  InfoBase Conference Papers}\ }\textbf {\bibinfo {volume} {6}},\ \bibinfo
  {pages} {829} (\bibinfo {year} {2012})}\BibitemShut {NoStop}%
\bibitem [{\citenamefont {Vincenti}(2019)}]{Vincenti2019}%
  \BibitemOpen
  \bibfield  {author} {\bibinfo {author} {\bibfnamefont {H.}~\bibnamefont
  {Vincenti}},\ }\href {\doibase 10.1103/PhysRevLett.123.105001} {\bibfield
  {journal} {\bibinfo  {journal} {Physical Review Letters}\ }\textbf {\bibinfo
  {volume} {123}},\ \bibinfo {pages} {105001} (\bibinfo {year} {2019})},\
  \Eprint {http://arxiv.org/abs/1812.05357} {arXiv:1812.05357} \BibitemShut
  {NoStop}%
\bibitem [{\citenamefont {Pirozhkov}\ \emph {et~al.}(2006)\citenamefont
  {Pirozhkov}, \citenamefont {Bulanov}, \citenamefont {Esirkepov},
  \citenamefont {Mori}, \citenamefont {Sagisaka},\ and\ \citenamefont
  {Daido}}]{Pirozhkov2006}%
  \BibitemOpen
  \bibfield  {author} {\bibinfo {author} {\bibfnamefont {A.~S.}\ \bibnamefont
  {Pirozhkov}}, \bibinfo {author} {\bibfnamefont {S.~V.}\ \bibnamefont
  {Bulanov}}, \bibinfo {author} {\bibfnamefont {T.~Z.}\ \bibnamefont
  {Esirkepov}}, \bibinfo {author} {\bibfnamefont {M.}~\bibnamefont {Mori}},
  \bibinfo {author} {\bibfnamefont {A.}~\bibnamefont {Sagisaka}}, \ and\
  \bibinfo {author} {\bibfnamefont {H.}~\bibnamefont {Daido}},\ }\href
  {\doibase 10.1063/1.2158145} {\bibfield  {journal} {\bibinfo  {journal}
  {Physics of Plasmas}\ }\textbf {\bibinfo {volume} {13}},\ \bibinfo {pages}
  {1} (\bibinfo {year} {2006})}\BibitemShut {NoStop}%
\bibitem [{\citenamefont {Pirozhkov}\ \emph
  {et~al.}(2007{\natexlab{a}})\citenamefont {Pirozhkov}, \citenamefont
  {Bulanov}, \citenamefont {Esirkepov}, \citenamefont {Sagisaka}, \citenamefont
  {Tajima},\ and\ \citenamefont {Daido}}]{Pirozhkov2007a}%
  \BibitemOpen
  \bibfield  {author} {\bibinfo {author} {\bibfnamefont {A.~S.}\ \bibnamefont
  {Pirozhkov}}, \bibinfo {author} {\bibfnamefont {S.~V.}\ \bibnamefont
  {Bulanov}}, \bibinfo {author} {\bibfnamefont {T.~Z.}\ \bibnamefont
  {Esirkepov}}, \bibinfo {author} {\bibfnamefont {A.}~\bibnamefont {Sagisaka}},
  \bibinfo {author} {\bibfnamefont {T.}~\bibnamefont {Tajima}}, \ and\ \bibinfo
  {author} {\bibfnamefont {H.}~\bibnamefont {Daido}},\ }\enquote {\bibinfo
  {title} {{Intensity scalings of attosecond pulse generation by the
  relativistic-irradiance laser pulses}},}\ in\ \href {\doibase
  10.1007/978-0-387-49119-6_35} {\emph {\bibinfo {booktitle} {Springer Series
  in Optical Sciences}}},\ Vol.\ \bibinfo {volume} {132},\ \bibinfo {editor}
  {edited by\ \bibinfo {editor} {\bibfnamefont {S.}~\bibnamefont {Watanabe}}\
  and\ \bibinfo {editor} {\bibfnamefont {K.}~\bibnamefont {Midorikawa}}}\
  (\bibinfo  {publisher} {Springer New York},\ \bibinfo {address} {New York,
  NY},\ \bibinfo {year} {2007})\ pp.\ \bibinfo {pages} {265--272}\BibitemShut
  {NoStop}%
\bibitem [{\citenamefont {Bulanov}, \citenamefont {Esirkepov},\ and\
  \citenamefont {Tajima}(2003)}]{Bulanov2003}%
  \BibitemOpen
  \bibfield  {author} {\bibinfo {author} {\bibfnamefont {S.~V.}\ \bibnamefont
  {Bulanov}}, \bibinfo {author} {\bibfnamefont {T.}~\bibnamefont {Esirkepov}},
  \ and\ \bibinfo {author} {\bibfnamefont {T.}~\bibnamefont {Tajima}},\ }\href
  {\doibase 10.1103/PhysRevLett.91.085001} {\bibfield  {journal} {\bibinfo
  {journal} {Physical Review Letters}\ }\textbf {\bibinfo {volume} {91}},\
  \bibinfo {pages} {85001} (\bibinfo {year} {2003})}\BibitemShut {NoStop}%
\bibitem [{\citenamefont {Kando}\ \emph {et~al.}(2007)\citenamefont {Kando},
  \citenamefont {Fukuda}, \citenamefont {Pirozhkov}, \citenamefont {Ma},
  \citenamefont {Daito}, \citenamefont {Chen}, \citenamefont {Esirkepov},
  \citenamefont {Ogura}, \citenamefont {Homma}, \citenamefont {Hayashi},
  \citenamefont {Kotaki}, \citenamefont {Sagisaka}, \citenamefont {Mori},
  \citenamefont {Koga}, \citenamefont {Daido}, \citenamefont {Bulanov},
  \citenamefont {Kimura}, \citenamefont {Kato},\ and\ \citenamefont
  {Tajima}}]{Kando2007}%
  \BibitemOpen
  \bibfield  {author} {\bibinfo {author} {\bibfnamefont {M.}~\bibnamefont
  {Kando}}, \bibinfo {author} {\bibfnamefont {Y.}~\bibnamefont {Fukuda}},
  \bibinfo {author} {\bibfnamefont {A.~S.}\ \bibnamefont {Pirozhkov}}, \bibinfo
  {author} {\bibfnamefont {J.}~\bibnamefont {Ma}}, \bibinfo {author}
  {\bibfnamefont {I.}~\bibnamefont {Daito}}, \bibinfo {author} {\bibfnamefont
  {L.~M.}\ \bibnamefont {Chen}}, \bibinfo {author} {\bibfnamefont {T.~Z.}\
  \bibnamefont {Esirkepov}}, \bibinfo {author} {\bibfnamefont {K.}~\bibnamefont
  {Ogura}}, \bibinfo {author} {\bibfnamefont {T.}~\bibnamefont {Homma}},
  \bibinfo {author} {\bibfnamefont {Y.}~\bibnamefont {Hayashi}}, \bibinfo
  {author} {\bibfnamefont {H.}~\bibnamefont {Kotaki}}, \bibinfo {author}
  {\bibfnamefont {A.}~\bibnamefont {Sagisaka}}, \bibinfo {author}
  {\bibfnamefont {M.}~\bibnamefont {Mori}}, \bibinfo {author} {\bibfnamefont
  {J.~K.}\ \bibnamefont {Koga}}, \bibinfo {author} {\bibfnamefont
  {H.}~\bibnamefont {Daido}}, \bibinfo {author} {\bibfnamefont {S.~V.}\
  \bibnamefont {Bulanov}}, \bibinfo {author} {\bibfnamefont {T.}~\bibnamefont
  {Kimura}}, \bibinfo {author} {\bibfnamefont {Y.}~\bibnamefont {Kato}}, \ and\
  \bibinfo {author} {\bibfnamefont {T.}~\bibnamefont {Tajima}},\ }\href
  {\doibase 10.1103/PhysRevLett.99.135001} {\bibfield  {journal} {\bibinfo
  {journal} {Physical Review Letters}\ }\textbf {\bibinfo {volume} {99}},\
  \bibinfo {pages} {135001} (\bibinfo {year} {2007})}\BibitemShut {NoStop}%
\bibitem [{\citenamefont {Pirozhkov}\ \emph
  {et~al.}(2007{\natexlab{b}})\citenamefont {Pirozhkov}, \citenamefont {Ma},
  \citenamefont {Kando}, \citenamefont {Esirkepov}, \citenamefont {Fukuda},
  \citenamefont {Chen}, \citenamefont {Daito}, \citenamefont {Ogura},
  \citenamefont {Homma}, \citenamefont {Hayashi}, \citenamefont {Kotaki},
  \citenamefont {Sagisaka}, \citenamefont {Mori}, \citenamefont {Koga},
  \citenamefont {Kawachi}, \citenamefont {Daido}, \citenamefont {Bulanov},
  \citenamefont {Kimura}, \citenamefont {Kato},\ and\ \citenamefont
  {Tajima}}]{Pirozhkov2007}%
  \BibitemOpen
  \bibfield  {author} {\bibinfo {author} {\bibfnamefont {A.~S.}\ \bibnamefont
  {Pirozhkov}}, \bibinfo {author} {\bibfnamefont {J.}~\bibnamefont {Ma}},
  \bibinfo {author} {\bibfnamefont {M.}~\bibnamefont {Kando}}, \bibinfo
  {author} {\bibfnamefont {T.~Z.}\ \bibnamefont {Esirkepov}}, \bibinfo {author}
  {\bibfnamefont {Y.}~\bibnamefont {Fukuda}}, \bibinfo {author} {\bibfnamefont
  {L.~M.}\ \bibnamefont {Chen}}, \bibinfo {author} {\bibfnamefont
  {I.}~\bibnamefont {Daito}}, \bibinfo {author} {\bibfnamefont
  {K.}~\bibnamefont {Ogura}}, \bibinfo {author} {\bibfnamefont
  {T.}~\bibnamefont {Homma}}, \bibinfo {author} {\bibfnamefont
  {Y.}~\bibnamefont {Hayashi}}, \bibinfo {author} {\bibfnamefont
  {H.}~\bibnamefont {Kotaki}}, \bibinfo {author} {\bibfnamefont
  {A.}~\bibnamefont {Sagisaka}}, \bibinfo {author} {\bibfnamefont
  {M.}~\bibnamefont {Mori}}, \bibinfo {author} {\bibfnamefont {J.~K.}\
  \bibnamefont {Koga}}, \bibinfo {author} {\bibfnamefont {T.}~\bibnamefont
  {Kawachi}}, \bibinfo {author} {\bibfnamefont {H.}~\bibnamefont {Daido}},
  \bibinfo {author} {\bibfnamefont {S.~V.}\ \bibnamefont {Bulanov}}, \bibinfo
  {author} {\bibfnamefont {T.}~\bibnamefont {Kimura}}, \bibinfo {author}
  {\bibfnamefont {Y.}~\bibnamefont {Kato}}, \ and\ \bibinfo {author}
  {\bibfnamefont {T.}~\bibnamefont {Tajima}},\ }\href {\doibase
  10.1063/1.2816443} {\bibfield  {journal} {\bibinfo  {journal} {Physics of
  Plasmas}\ }\textbf {\bibinfo {volume} {14}},\ \bibinfo {pages} {123106}
  (\bibinfo {year} {2007}{\natexlab{b}})}\BibitemShut {NoStop}%
\bibitem [{\citenamefont {Kando}\ \emph {et~al.}(2009)\citenamefont {Kando},
  \citenamefont {Pirozhkov}, \citenamefont {Kawase}, \citenamefont {Esirkepov},
  \citenamefont {Fukuda}, \citenamefont {Kiriyama}, \citenamefont {Okada},
  \citenamefont {Daito}, \citenamefont {Kameshima}, \citenamefont {Hayashi},
  \citenamefont {Kotaki}, \citenamefont {Mori}, \citenamefont {Koga},
  \citenamefont {Daido}, \citenamefont {Faenov}, \citenamefont {Pikuz},
  \citenamefont {Ma}, \citenamefont {Chen}, \citenamefont {Ragozin},
  \citenamefont {Kawachi}, \citenamefont {Kato}, \citenamefont {Tajima},\ and\
  \citenamefont {Bulanov}}]{Kando2009}%
  \BibitemOpen
  \bibfield  {author} {\bibinfo {author} {\bibfnamefont {M.}~\bibnamefont
  {Kando}}, \bibinfo {author} {\bibfnamefont {A.~S.}\ \bibnamefont
  {Pirozhkov}}, \bibinfo {author} {\bibfnamefont {K.}~\bibnamefont {Kawase}},
  \bibinfo {author} {\bibfnamefont {T.~Z.}\ \bibnamefont {Esirkepov}}, \bibinfo
  {author} {\bibfnamefont {Y.}~\bibnamefont {Fukuda}}, \bibinfo {author}
  {\bibfnamefont {H.}~\bibnamefont {Kiriyama}}, \bibinfo {author}
  {\bibfnamefont {H.}~\bibnamefont {Okada}}, \bibinfo {author} {\bibfnamefont
  {I.}~\bibnamefont {Daito}}, \bibinfo {author} {\bibfnamefont
  {T.}~\bibnamefont {Kameshima}}, \bibinfo {author} {\bibfnamefont
  {Y.}~\bibnamefont {Hayashi}}, \bibinfo {author} {\bibfnamefont
  {H.}~\bibnamefont {Kotaki}}, \bibinfo {author} {\bibfnamefont
  {M.}~\bibnamefont {Mori}}, \bibinfo {author} {\bibfnamefont {J.~K.}\
  \bibnamefont {Koga}}, \bibinfo {author} {\bibfnamefont {H.}~\bibnamefont
  {Daido}}, \bibinfo {author} {\bibfnamefont {A.~Y.}\ \bibnamefont {Faenov}},
  \bibinfo {author} {\bibfnamefont {T.}~\bibnamefont {Pikuz}}, \bibinfo
  {author} {\bibfnamefont {J.}~\bibnamefont {Ma}}, \bibinfo {author}
  {\bibfnamefont {L.~M.}\ \bibnamefont {Chen}}, \bibinfo {author}
  {\bibfnamefont {E.~N.}\ \bibnamefont {Ragozin}}, \bibinfo {author}
  {\bibfnamefont {T.}~\bibnamefont {Kawachi}}, \bibinfo {author} {\bibfnamefont
  {Y.}~\bibnamefont {Kato}}, \bibinfo {author} {\bibfnamefont {T.}~\bibnamefont
  {Tajima}}, \ and\ \bibinfo {author} {\bibfnamefont {S.~V.}\ \bibnamefont
  {Bulanov}},\ }\href {\doibase 10.1103/PhysRevLett.103.235003} {\bibfield
  {journal} {\bibinfo  {journal} {Physical Review Letters}\ }\textbf {\bibinfo
  {volume} {103}},\ \bibinfo {pages} {235003} (\bibinfo {year}
  {2009})}\BibitemShut {NoStop}%
\bibitem [{\citenamefont {Lobet}\ \emph {et~al.}(2013)\citenamefont {Lobet},
  \citenamefont {Kando}, \citenamefont {Koga}, \citenamefont {Esirkepov},
  \citenamefont {Nakamura}, \citenamefont {Pirozhkov},\ and\ \citenamefont
  {Bulanov}}]{Lobet2013}%
  \BibitemOpen
  \bibfield  {author} {\bibinfo {author} {\bibfnamefont {M.}~\bibnamefont
  {Lobet}}, \bibinfo {author} {\bibfnamefont {M.}~\bibnamefont {Kando}},
  \bibinfo {author} {\bibfnamefont {J.~K.}\ \bibnamefont {Koga}}, \bibinfo
  {author} {\bibfnamefont {T.~Z.}\ \bibnamefont {Esirkepov}}, \bibinfo {author}
  {\bibfnamefont {T.}~\bibnamefont {Nakamura}}, \bibinfo {author}
  {\bibfnamefont {A.~S.}\ \bibnamefont {Pirozhkov}}, \ and\ \bibinfo {author}
  {\bibfnamefont {S.~V.}\ \bibnamefont {Bulanov}},\ }\href {\doibase
  10.1016/j.physleta.2013.02.042} {\bibfield  {journal} {\bibinfo  {journal}
  {Physics Letters, Section A: General, Atomic and Solid State Physics}\
  }\textbf {\bibinfo {volume} {377}},\ \bibinfo {pages} {1114} (\bibinfo {year}
  {2013})}\BibitemShut {NoStop}%
\bibitem [{\citenamefont {Koga}\ \emph {et~al.}(2018)\citenamefont {Koga},
  \citenamefont {Bulanov}, \citenamefont {Esirkepov}, \citenamefont {Kando},
  \citenamefont {Bulanov},\ and\ \citenamefont {Pirozhkov}}]{Koga2018}%
  \BibitemOpen
  \bibfield  {author} {\bibinfo {author} {\bibfnamefont {J.~K.}\ \bibnamefont
  {Koga}}, \bibinfo {author} {\bibfnamefont {S.~V.}\ \bibnamefont {Bulanov}},
  \bibinfo {author} {\bibfnamefont {T.~Z.}\ \bibnamefont {Esirkepov}}, \bibinfo
  {author} {\bibfnamefont {M.}~\bibnamefont {Kando}}, \bibinfo {author}
  {\bibfnamefont {S.~S.}\ \bibnamefont {Bulanov}}, \ and\ \bibinfo {author}
  {\bibfnamefont {A.~S.}\ \bibnamefont {Pirozhkov}},\ }\href {\doibase
  10.1088/1361-6587/aac068} {\bibfield  {journal} {\bibinfo  {journal} {Plasma
  Physics and Controlled Fusion}\ }\textbf {\bibinfo {volume} {60}},\ \bibinfo
  {pages} {74007} (\bibinfo {year} {2018})}\BibitemShut {NoStop}%
\bibitem [{\citenamefont {Moghadasin}\ \emph {et~al.}(2019)\citenamefont
  {Moghadasin}, \citenamefont {Niknam}, \citenamefont {Komaizi},\ and\
  \citenamefont {Banjafar}}]{Moghadasin2019}%
  \BibitemOpen
  \bibfield  {author} {\bibinfo {author} {\bibfnamefont {H.}~\bibnamefont
  {Moghadasin}}, \bibinfo {author} {\bibfnamefont {A.~R.}\ \bibnamefont
  {Niknam}}, \bibinfo {author} {\bibfnamefont {D.}~\bibnamefont {Komaizi}}, \
  and\ \bibinfo {author} {\bibfnamefont {M.}~\bibnamefont {Banjafar}},\ }\href
  {\doibase 10.1063/1.5119041} {\bibfield  {journal} {\bibinfo  {journal}
  {Physics of Plasmas}\ }\textbf {\bibinfo {volume} {26}},\ \bibinfo {pages}
  {93105} (\bibinfo {year} {2019})}\BibitemShut {NoStop}%
\bibitem [{\citenamefont {Mu}\ \emph {et~al.}(2019)\citenamefont {Mu},
  \citenamefont {Esirkepov}, \citenamefont {Valenta}, \citenamefont {Jeong},
  \citenamefont {Gu}, \citenamefont {Koga}, \citenamefont {Pirozhkov},
  \citenamefont {Kando}, \citenamefont {Korn},\ and\ \citenamefont
  {Bulanov}}]{Mu2019}%
  \BibitemOpen
  \bibfield  {author} {\bibinfo {author} {\bibfnamefont {J.}~\bibnamefont
  {Mu}}, \bibinfo {author} {\bibfnamefont {T.~Z.}\ \bibnamefont {Esirkepov}},
  \bibinfo {author} {\bibfnamefont {P.}~\bibnamefont {Valenta}}, \bibinfo
  {author} {\bibfnamefont {T.~M.}\ \bibnamefont {Jeong}}, \bibinfo {author}
  {\bibfnamefont {Y.}~\bibnamefont {Gu}}, \bibinfo {author} {\bibfnamefont
  {J.~K.}\ \bibnamefont {Koga}}, \bibinfo {author} {\bibfnamefont {A.~S.}\
  \bibnamefont {Pirozhkov}}, \bibinfo {author} {\bibfnamefont {M.}~\bibnamefont
  {Kando}}, \bibinfo {author} {\bibfnamefont {G.}~\bibnamefont {Korn}}, \ and\
  \bibinfo {author} {\bibfnamefont {S.~V.}\ \bibnamefont {Bulanov}},\ }\href
  {\doibase 10.3103/S1541308X19040010} {\bibfield  {journal} {\bibinfo
  {journal} {Physics of Wave Phenomena}\ }\textbf {\bibinfo {volume} {27}},\
  \bibinfo {pages} {247} (\bibinfo {year} {2019})}\BibitemShut {NoStop}%
\bibitem [{\citenamefont {Koga}\ \emph {et~al.}(2012)\citenamefont {Koga},
  \citenamefont {Bulanov}, \citenamefont {Esirkepov}, \citenamefont
  {Pirozhkov}, \citenamefont {Kando},\ and\ \citenamefont
  {Rosanov}}]{Koga2012}%
  \BibitemOpen
  \bibfield  {author} {\bibinfo {author} {\bibfnamefont {J.~K.}\ \bibnamefont
  {Koga}}, \bibinfo {author} {\bibfnamefont {S.~V.}\ \bibnamefont {Bulanov}},
  \bibinfo {author} {\bibfnamefont {T.~Z.}\ \bibnamefont {Esirkepov}}, \bibinfo
  {author} {\bibfnamefont {A.~S.}\ \bibnamefont {Pirozhkov}}, \bibinfo {author}
  {\bibfnamefont {M.}~\bibnamefont {Kando}}, \ and\ \bibinfo {author}
  {\bibfnamefont {N.~N.}\ \bibnamefont {Rosanov}},\ }\href {\doibase
  10.1103/PhysRevA.86.053823} {\bibfield  {journal} {\bibinfo  {journal} {Phys.
  Rev. A}\ }\textbf {\bibinfo {volume} {86}},\ \bibinfo {pages} {53823}
  (\bibinfo {year} {2012})}\BibitemShut {NoStop}%
\bibitem [{\citenamefont {Chen}\ and\ \citenamefont {Mourou}(2017)}]{Chen2017}%
  \BibitemOpen
  \bibfield  {author} {\bibinfo {author} {\bibfnamefont {P.}~\bibnamefont
  {Chen}}\ and\ \bibinfo {author} {\bibfnamefont {G.}~\bibnamefont {Mourou}},\
  }\href {\doibase 10.1103/PhysRevLett.118.045001} {\bibfield  {journal}
  {\bibinfo  {journal} {Physical Review Letters}\ }\textbf {\bibinfo {volume}
  {118}},\ \bibinfo {pages} {45001} (\bibinfo {year} {2017})},\ \Eprint
  {http://arxiv.org/abs/1512.04064} {arXiv:1512.04064} \BibitemShut {NoStop}%
\bibitem [{\citenamefont {Bulanov}\ \emph {et~al.}(2002)\citenamefont
  {Bulanov}, \citenamefont {Esirkepov}, \citenamefont {Khoroshkov},
  \citenamefont {Kuznetsov},\ and\ \citenamefont {Pegoraro}}]{Bulanov2002}%
  \BibitemOpen
  \bibfield  {author} {\bibinfo {author} {\bibfnamefont {S.~V.}\ \bibnamefont
  {Bulanov}}, \bibinfo {author} {\bibfnamefont {T.}~\bibnamefont {Esirkepov}},
  \bibinfo {author} {\bibfnamefont {V.~S.}\ \bibnamefont {Khoroshkov}},
  \bibinfo {author} {\bibfnamefont {A.~V.}\ \bibnamefont {Kuznetsov}}, \ and\
  \bibinfo {author} {\bibfnamefont {F.}~\bibnamefont {Pegoraro}},\ }\href
  {\doibase 10.1016/S0375-9601(02)00521-2} {\bibfield  {journal} {\bibinfo
  {journal} {Physics Letters, Section A: General, Atomic and Solid State
  Physics}\ }\textbf {\bibinfo {volume} {299}},\ \bibinfo {pages} {240}
  (\bibinfo {year} {2002})}\BibitemShut {NoStop}%
\bibitem [{\citenamefont {Neutze}\ \emph {et~al.}(2000)\citenamefont {Neutze},
  \citenamefont {Wouts}, \citenamefont {{Van Der Spoel}}, \citenamefont
  {Weckert},\ and\ \citenamefont {Hajdu}}]{Neutze2000}%
  \BibitemOpen
  \bibfield  {author} {\bibinfo {author} {\bibfnamefont {R.}~\bibnamefont
  {Neutze}}, \bibinfo {author} {\bibfnamefont {R.}~\bibnamefont {Wouts}},
  \bibinfo {author} {\bibfnamefont {D.}~\bibnamefont {{Van Der Spoel}}},
  \bibinfo {author} {\bibfnamefont {E.}~\bibnamefont {Weckert}}, \ and\
  \bibinfo {author} {\bibfnamefont {J.}~\bibnamefont {Hajdu}},\ }\href
  {\doibase 10.1038/35021099} {\bibfield  {journal} {\bibinfo  {journal}
  {Nature}\ }\textbf {\bibinfo {volume} {406}},\ \bibinfo {pages} {752}
  (\bibinfo {year} {2000})}\BibitemShut {NoStop}%
\bibitem [{\citenamefont {Krausz}\ and\ \citenamefont
  {Ivanov}(2009)}]{Krausz2009}%
  \BibitemOpen
  \bibfield  {author} {\bibinfo {author} {\bibfnamefont {F.}~\bibnamefont
  {Krausz}}\ and\ \bibinfo {author} {\bibfnamefont {M.}~\bibnamefont
  {Ivanov}},\ }\href {\doibase 10.1103/RevModPhys.81.163} {\bibfield  {journal}
  {\bibinfo  {journal} {Rev. Mod. Phys.}\ }\textbf {\bibinfo {volume} {81}},\
  \bibinfo {pages} {163} (\bibinfo {year} {2009})}\BibitemShut {NoStop}%
\bibitem [{\citenamefont {Bulanov}\ \emph {et~al.}(2016)\citenamefont
  {Bulanov}, \citenamefont {Esirkepov}, \citenamefont {Hayashi}, \citenamefont
  {Kiriyama}, \citenamefont {Koga}, \citenamefont {Kotaki}, \citenamefont
  {Mori},\ and\ \citenamefont {Kando}}]{Bulanov2016}%
  \BibitemOpen
  \bibfield  {author} {\bibinfo {author} {\bibfnamefont {S.~V.}\ \bibnamefont
  {Bulanov}}, \bibinfo {author} {\bibfnamefont {T.~Z.}\ \bibnamefont
  {Esirkepov}}, \bibinfo {author} {\bibfnamefont {Y.}~\bibnamefont {Hayashi}},
  \bibinfo {author} {\bibfnamefont {H.}~\bibnamefont {Kiriyama}}, \bibinfo
  {author} {\bibfnamefont {J.~K.}\ \bibnamefont {Koga}}, \bibinfo {author}
  {\bibfnamefont {H.}~\bibnamefont {Kotaki}}, \bibinfo {author} {\bibfnamefont
  {M.}~\bibnamefont {Mori}}, \ and\ \bibinfo {author} {\bibfnamefont
  {M.}~\bibnamefont {Kando}},\ }\href {\doibase 10.1017/S0022377816000623}
  {\bibfield  {journal} {\bibinfo  {journal} {Journal of Plasma Physics}\
  }\textbf {\bibinfo {volume} {82}},\ \bibinfo {pages} {905820308} (\bibinfo
  {year} {2016})}\BibitemShut {NoStop}%
\bibitem [{\citenamefont {Esarey}, \citenamefont {Schroeder},\ and\
  \citenamefont {Leemans}(2009)}]{Esarey2009}%
  \BibitemOpen
  \bibfield  {author} {\bibinfo {author} {\bibfnamefont {E.}~\bibnamefont
  {Esarey}}, \bibinfo {author} {\bibfnamefont {C.~B.}\ \bibnamefont
  {Schroeder}}, \ and\ \bibinfo {author} {\bibfnamefont {W.~P.}\ \bibnamefont
  {Leemans}},\ }\href {\doibase 10.1103/RevModPhys.81.1229} {\bibfield
  {journal} {\bibinfo  {journal} {Reviews of Modern Physics}\ }\textbf
  {\bibinfo {volume} {81}},\ \bibinfo {pages} {1229} (\bibinfo {year}
  {2009})}\BibitemShut {NoStop}%
\bibitem [{\citenamefont {Akhiezer}\ and\ \citenamefont
  {Polovin}(1956)}]{Akhiezer1956}%
  \BibitemOpen
  \bibfield  {author} {\bibinfo {author} {\bibfnamefont {A.}~\bibnamefont
  {Akhiezer}}\ and\ \bibinfo {author} {\bibfnamefont {R.}~\bibnamefont
  {Polovin}},\ }\href@noop {} {\bibfield  {journal} {\bibinfo  {journal}
  {Soviet Phys. JETP}\ }\textbf {\bibinfo {volume} {Vol: 3}},\ \bibinfo {pages}
  {696} (\bibinfo {year} {1956})}\BibitemShut {NoStop}%
\bibitem [{\citenamefont {Abramowitz}\ and\ \citenamefont
  {Stegun}(1965)}]{Abramowitz1965}%
  \BibitemOpen
  \bibfield  {author} {\bibinfo {author} {\bibfnamefont {M.}~\bibnamefont
  {Abramowitz}}\ and\ \bibinfo {author} {\bibfnamefont {I.~A.}\ \bibnamefont
  {Stegun}},\ }\href {\doibase 10.2307/2343473} {\emph {\bibinfo {title}
  {Journal of the Royal Statistical Society. Series A (General)}}},\ \bibinfo
  {series} {Applied mathematics series}, Vol.\ \bibinfo {volume} {128}\
  (\bibinfo  {publisher} {Dover Publications},\ \bibinfo {year} {1965})\ p.\
  \bibinfo {pages} {593}\BibitemShut {NoStop}%
\bibitem [{\citenamefont {Arber}\ \emph {et~al.}(2015)\citenamefont {Arber},
  \citenamefont {Bennett}, \citenamefont {Brady}, \citenamefont
  {Lawrence-Douglas}, \citenamefont {Ramsay}, \citenamefont {Sircombe},
  \citenamefont {Gillies}, \citenamefont {Evans}, \citenamefont {Schmitz},
  \citenamefont {Bell},\ and\ \citenamefont {Ridgers}}]{Arber2015}%
  \BibitemOpen
  \bibfield  {author} {\bibinfo {author} {\bibfnamefont {T.~D.}\ \bibnamefont
  {Arber}}, \bibinfo {author} {\bibfnamefont {K.}~\bibnamefont {Bennett}},
  \bibinfo {author} {\bibfnamefont {C.~S.}\ \bibnamefont {Brady}}, \bibinfo
  {author} {\bibfnamefont {A.}~\bibnamefont {Lawrence-Douglas}}, \bibinfo
  {author} {\bibfnamefont {M.~G.}\ \bibnamefont {Ramsay}}, \bibinfo {author}
  {\bibfnamefont {N.~J.}\ \bibnamefont {Sircombe}}, \bibinfo {author}
  {\bibfnamefont {P.}~\bibnamefont {Gillies}}, \bibinfo {author} {\bibfnamefont
  {R.~G.}\ \bibnamefont {Evans}}, \bibinfo {author} {\bibfnamefont
  {H.}~\bibnamefont {Schmitz}}, \bibinfo {author} {\bibfnamefont {A.~R.}\
  \bibnamefont {Bell}}, \ and\ \bibinfo {author} {\bibfnamefont {C.~P.}\
  \bibnamefont {Ridgers}},\ }\href {\doibase 10.1088/0741-3335/57/11/113001}
  {\bibfield  {journal} {\bibinfo  {journal} {Plasma Physics and Controlled
  Fusion}\ }\textbf {\bibinfo {volume} {57}},\ \bibinfo {pages} {113001}
  (\bibinfo {year} {2015})}\BibitemShut {NoStop}%
\bibitem [{\citenamefont {Bulanov}\ \emph {et~al.}(1990)\citenamefont
  {Bulanov}, \citenamefont {Inovenkov}, \citenamefont {Naumova},\ and\
  \citenamefont {Sakharov}}]{Bulanov1990}%
  \BibitemOpen
  \bibfield  {author} {\bibinfo {author} {\bibfnamefont {S.~V.}\ \bibnamefont
  {Bulanov}}, \bibinfo {author} {\bibfnamefont {I.~N.}\ \bibnamefont
  {Inovenkov}}, \bibinfo {author} {\bibfnamefont {N.~M.}\ \bibnamefont
  {Naumova}}, \ and\ \bibinfo {author} {\bibfnamefont {A.~S.}\ \bibnamefont
  {Sakharov}},\ }\href@noop {} {\bibfield  {journal} {\bibinfo  {journal} {Sov.
  J. Plasma Phys.}\ }\textbf {\bibinfo {volume} {16}},\ \bibinfo {pages} {444}
  (\bibinfo {year} {1990})}\BibitemShut {NoStop}%
\bibitem [{\citenamefont {Fidel}\ \emph {et~al.}(1997)\citenamefont {Fidel},
  \citenamefont {Heyman}, \citenamefont {Kastner},\ and\ \citenamefont
  {Ziolkowski}}]{Fidel1997}%
  \BibitemOpen
  \bibfield  {author} {\bibinfo {author} {\bibfnamefont {B.}~\bibnamefont
  {Fidel}}, \bibinfo {author} {\bibfnamefont {E.}~\bibnamefont {Heyman}},
  \bibinfo {author} {\bibfnamefont {R.}~\bibnamefont {Kastner}}, \ and\
  \bibinfo {author} {\bibfnamefont {R.~W.}\ \bibnamefont {Ziolkowski}},\ }\href
  {\doibase 10.1006/jcph.1997.5827} {\bibfield  {journal} {\bibinfo  {journal}
  {Journal of Computational Physics}\ }\textbf {\bibinfo {volume} {138}},\
  \bibinfo {pages} {480} (\bibinfo {year} {1997})}\BibitemShut {NoStop}%
\bibitem [{\citenamefont {Yee}(1966)}]{Yee1966}%
  \BibitemOpen
  \bibfield  {author} {\bibinfo {author} {\bibfnamefont {K.~S.}\ \bibnamefont
  {Yee}},\ }\href {\doibase 10.1109/TAP.1966.1138693} {\bibfield  {journal}
  {\bibinfo  {journal} {IEEE Transactions on Antennas and Propagation}\
  }\textbf {\bibinfo {volume} {14}},\ \bibinfo {pages} {302} (\bibinfo {year}
  {1966})}\BibitemShut {NoStop}%
\bibitem [{\citenamefont {Courant}, \citenamefont {Friedrichs},\ and\
  \citenamefont {Lewy}(1928)}]{Courant1928}%
  \BibitemOpen
  \bibfield  {author} {\bibinfo {author} {\bibfnamefont {R.}~\bibnamefont
  {Courant}}, \bibinfo {author} {\bibfnamefont {K.}~\bibnamefont {Friedrichs}},
  \ and\ \bibinfo {author} {\bibfnamefont {H.}~\bibnamefont {Lewy}},\ }\href
  {\doibase 10.1007/BF01448839} {\bibfield  {journal} {\bibinfo  {journal}
  {Mathematische Annalen}\ }\textbf {\bibinfo {volume} {100}},\ \bibinfo
  {pages} {32} (\bibinfo {year} {1928})}\BibitemShut {NoStop}%
\bibitem [{\citenamefont {Savitzky}\ and\ \citenamefont
  {Golay}(1964)}]{Savitzky1964}%
  \BibitemOpen
  \bibfield  {author} {\bibinfo {author} {\bibfnamefont {A.}~\bibnamefont
  {Savitzky}}\ and\ \bibinfo {author} {\bibfnamefont {M.~J.}\ \bibnamefont
  {Golay}},\ }\href {\doibase 10.1021/ac60214a047} {\bibfield  {journal}
  {\bibinfo  {journal} {Analytical Chemistry}\ }\textbf {\bibinfo {volume}
  {36}},\ \bibinfo {pages} {1627} (\bibinfo {year} {1964})}\BibitemShut
  {NoStop}%
\bibitem [{\citenamefont {Bulanov}\ \emph {et~al.}(1992)\citenamefont
  {Bulanov}, \citenamefont {Inovenkov}, \citenamefont {Kirsanov}, \citenamefont
  {Naumova},\ and\ \citenamefont {Sakharov}}]{Bulanov1992}%
  \BibitemOpen
  \bibfield  {author} {\bibinfo {author} {\bibfnamefont {S.~V.}\ \bibnamefont
  {Bulanov}}, \bibinfo {author} {\bibfnamefont {I.~N.}\ \bibnamefont
  {Inovenkov}}, \bibinfo {author} {\bibfnamefont {V.~I.}\ \bibnamefont
  {Kirsanov}}, \bibinfo {author} {\bibfnamefont {N.~M.}\ \bibnamefont
  {Naumova}}, \ and\ \bibinfo {author} {\bibfnamefont {A.~S.}\ \bibnamefont
  {Sakharov}},\ }\href {\doibase 10.1063/1.860046} {\bibfield  {journal}
  {\bibinfo  {journal} {Physics of Fluids B}\ }\textbf {\bibinfo {volume}
  {4}},\ \bibinfo {pages} {1935} (\bibinfo {year} {1992})}\BibitemShut
  {NoStop}%
\bibitem [{\citenamefont {King}(2009)}]{King2009}%
  \BibitemOpen
  \bibfield  {author} {\bibinfo {author} {\bibfnamefont {F.~W.}\ \bibnamefont
  {King}},\ }\href {\doibase 10.1017/cbo9780511735271} {\emph {\bibinfo {title}
  {Hilbert transforms}}},\ \bibinfo {series} {Encyclopedia of Mathematics and
  its Applications}, Vol.~\bibinfo {volume} {2}\ (\bibinfo  {publisher}
  {Cambridge University Press},\ \bibinfo {year} {2009})\BibitemShut {NoStop}%
\end{thebibliography}%

\end{document}